\documentclass[%
 reprint,
superscriptaddress,
 amsmath,amssymb,
 aps,
prl,
]{revtex4-1}

\usepackage{comment}
\usepackage{graphicx}
\usepackage{dcolumn}
\usepackage{bm}

\begin{document}


\title{Accelerating Science with Generative Adversarial Networks:\\An Application to 3D Particle Showers in Multi-Layer Calorimeters}

\author{Michela Paganini}
\email{michela.paganini@yale.edu}
\affiliation{Lawrence Berkeley National Laboratory, Berkeley, CA 94720}
\affiliation{Yale University, New Haven, CT 06520}

\author{Luke de Oliveira}
\email{lukedeoliveira@lbl.gov}
\affiliation{Lawrence Berkeley National Laboratory, Berkeley, CA 94720}

\author{Benjamin Nachman}
\email{bnachman@cern.ch}
\affiliation{Lawrence Berkeley National Laboratory, Berkeley, CA 94720}



\begin{abstract}

Physicists at the Large Hadron Collider (LHC) rely on detailed simulations of particle collisions to build expectations of what experimental data may look like under different theory modeling assumptions. Petabytes of simulated data are needed to develop analysis techniques, though they are expensive to generate using existing algorithms and computing resources. The modeling of detectors and the precise description of particle cascades as they interact with the material in the calorimeter are the most computationally demanding steps in the simulation pipeline. 
We therefore introduce a deep neural network-based generative model to enable high-fidelity, fast, electromagnetic calorimeter simulation. There are still challenges for achieving precision across the entire phase space, but our current solution can reproduce a variety of particle shower properties while achieving speed-up factors of up to 100,000$\times$. This opens the door to a new era of fast simulation that could save significant computing time and disk space, while extending the reach of physics searches and precision measurements at the LHC and beyond.

\end{abstract}

\maketitle


%

\section{Introduction}
\label{sec:intro}
High-precision modeling of the interactions of particles with media is important across many physical sciences, enabling and accelerating new findings. Similar to complex weather or cosmological modeling, the detailed simulation of subatomic particle collisions and interactions, as captured by detectors at the LHC, is a computationally demanding task, which annually requires billions of CPU hours, constituting more than half of the LHC experiments' computing resources~\cite{Flynn:2002240,dashboard,Bozzi:1984010}.

The Nobel-prize-winning Higgs boson discovery~\cite{Aad:2012tfa,Chatrchyan:2012xdj} would not have been possible without extensive simulation. Before its experimental observation, its fundamental properties, such as its mass, were unknown, but synthetic particle collisions could be generated to simulate the outcome of various measurements under different model assumptions.

Today, as several questions remain unanswered about the nature of known particles (such as neutrinos) and hypothetical ones (such as the supersymmetric partners of the Standard Model particles), 
modern nuclear and particle physics research continues to strongly depend on detailed simulations for developing analysis techniques, interpreting results, and designing new experiments.

Cutting-edge software libraries such as \textsc{Geant4}~\cite{Geant} provide the backbone to construct complex detector geometries and accurately model physical processes and interactions happening at distance scales as small as $10^{-20}$ m. 

The shortcoming of this method is its computational footprint. The high-precision description of electromagnetic and nuclear processes that govern the evolution of particle showers in calorimeters can requires minutes per event on modern computing platforms~\cite{Aad:2010ah,1742-6596-396-6-062016}, making this the most computationally expensive step in the simulation pipeline. Due to the expensive simulation cost, significant resources are also invested in storing generated data sets, which can occupy petabytes of disk space.

This bottleneck becomes apparent at the scale at which events need to be simulated to enable physics analyses at the high luminosity phase of the LHC (HL-LHC). The ATLAS and CMS experiments are expected to observe about $10^8$ Higgs boson events~\cite{deFlorian:2016spz}, buried in $\sim 10^{17}$ background events~\cite{Aaboud:2016mmw,CMS-PAS-FSQ-15-005}. Hundreds of billions of simulated collisions will be required to reduce the Monte Carlo uncertainty and measure some of the Higgs boson's as yet unprobed properties.

Approximate calorimeter simulation techniques exist~\cite{Grindhammer:1993kw,ATLAS:1300517,Grindhammer:1989zg, frozen}, but they provide compromises that lie on different, yet similarly sub-optimal, parts of the accuracy-speedup trade-off curve. 

Full detector simulations are too slow to meet the growing analysis demands; current fast simulations are not precise enough to serve the entire physics program. We therefore introduce a Deep Learning model, named \textsc{CaloGAN}, for high-fidelity fast simulation of particle showers in electromagnetic calorimeters. Its goal is to be both quick and precise, by significantly reducing the accuracy cost incurred with increased speed-up. A fast simulation technique of this kind also addresses the issue of data storage and transfer, as the gained generation simplicity and speedup make real-time, on-demand simulation a possibility. 

Similar techniques have been tested in Cosmology~\cite{CosmoGAN1,CosmoGAN2}, Condensed Matter Physics~\cite{condmatter}, and Oncology~\cite{oncology}. However, the sparsity, high dynamic range, and highly location-dependent features present in this application make it uniquely challenging.
In addition to enabling physics analysis at the LHC, an approach similar to the \textsc{CaloGAN} may be useful for other applications in particle and nuclear physics, nuclear medicine, and space science that require detailed modeling of particle interactions with matter.

\section{Method}
\label{sec:method}

To alleviate the computational burden of simulating electromagnetic showers, we introduce a method based on Generative Adversarial Networks (GANs) \cite{goodfellow2014generative} in order to directly simulate component read-outs in electromagnetic calorimeters. GANs are an increasingly popular approach to learning a generative model using deep neural networks, and have shown great promise in generating clear samples from natural images~\cite{dcgan}. 

Though the GAN formulation, by design, does not admit an explicit probability density or explicit likelihood, we gain the ability to sample from the learned generative model in a efficient manner. The GAN training uses a minimax game theoretic framework, and admits a function $g$ as an artifact that maps a $d$-dimensional latent vector, $z\sim p_z(z) \in\mathbb{R}^{d}$ to a point in the space of realistic samples. We would like the implicit density learned by $g$ to be close to the distribution $f$ that governs the simulated data distribution. Since $g$ is a neural network, a forward pass to generate new samples is highly efficient on modern computing platforms~\cite{DBLP:journals/corr/ChetlurWVCTCS14}.



Previous work~\cite{deOliveira:2017pjk} investigated GAN-based methods for jet images~\cite{Cogan:2014oua}, which are similar to one-layer calorimeters with square pixels (except jet generatators such as \textsc{Pythia}~\cite{Sjostrand:2006za} are much faster than \textsc{Geant4}). This work addresses the complexity introduced by modeling a realistic sampling detector with heterogeneous longitudinal and transverse segmentation.
We exploit the location specificity of the calorimeter, and utilize weight locality at the model level. We also follow the guidelines outlined in~\cite{deOliveira:2017pjk} in order to deal with both high dynamic range and sparsity levels. Our neural network architecture per calorimeter layer is a function of the read-out grid dimensionality, and is augmented with an attentional component~\cite{attention} that provides a mechanism to carry information from layer to layer~\cite{stackgan}. This allows the \textsc{CaloGAN} to model the physical sequential dependence among the calorimeter layers.

To ensure the realism of the \textsc{CaloGAN} setup, we impose an additional constraint to encourage the generator to produce a given energy shower. That is, the learned, implicit PDF $f$ needs to converge to the hypothetical data generating function $g$ for any initial nominal energy $E_0$, i.e., that $f(x \vert E=E_0)\longrightarrow g(x \vert E=E_0)$ for all $E_0\in[E_{\mathsf{min}}, E_{\mathsf{max}}]$. 

To encourage this to be well modeled, a physics-specific loss component is introduced to penalize absolute deviation between the nominal energy $E_0$ and the reconstructed energy $\widehat{E}$. A noteworthy subtlety is that this penalization scheme, coupled with minibatch discrimination~\cite{improved_gan}, invites the network to learn the distribution of $\vert E_0 - \widehat{E}\vert$, a desirable characteristic for a readily applicable practical system to augment fast simulation. Such a formulation also encourages conservation of energy through the generation process.  The simulation only includes models of energy deposition, not digitization (a non-linear effect that can violate reconstructed energy conservation).  The energy per layer includes the contribution from inactive material (see below).  Therefore, aside from leakage beyond the calorimeter (relevant mostly for charged pions), energy must be conserved and provides a useful constraint on the generation.

\section{Experimental Results}
\label{sec:expresults}
From a series of simulated showers, the \textsc{CaloGAN} is tasked with learning the simulated data distributions of $\gamma$, $e^+$, and $\pi^+$ generated by \textsc{Geant4} with uniform energy spectrum $[1, 100]$ GeV, and incident perpendicular to the center of a three-layer, heterogeneously segmented, liquid argon (LAr) calorimeter cube of side-length $480$ mm. The training dataset~\cite{dataset} is represented in image format by three figures of dimensions $3\times 96$, $12\times 12$, and $12\times 6$, each representing the shower energy depositions per pixel in each calorimeter layer. The energy per layer includes the active and inactive contributions. For e.g. calorimeter calibrations~\cite{Aad:2016upy}, it is important to have the inactive component; in the future one could add separate layers for the inactive component or add a second step for dividing the energy per layer into the two components. The flexible \textsc{CaloGAN} architecture allows for a straightforward extension to related detector geometries that have more sampling layers or different cell sizes per layer~\cite{code_new}. 

Our analysis establishes that it is possible to generate three-dimensional electromagnetic showers in a multi-layer sampling LAr calorimeter with uneven spatial segmentation, while attempting to preserve spatio-temporal relationships among layers. 

For performance evaluation, we choose application-driven methods focused on sample quality. A first qualitative assessment is accompanied by a quantitative evaluation based on physics-driven similarity metrics. The choice reflects the domain specific procedure for Monte Carlo-data comparisons.  However, it is also important to examine high-dimensional behavior because \textsc{CaloGAN} is not anchored by parameterized models the way traditional fast simulators are.  While the adversarial classifier provides some high-dimensional validation, we also use particle classification performance. Visualization and validation is still a key challenge for multi-dimensional generators parameterized by a neural network.

\subsection{Qualitative Evaluation}

\begin{figure}[h!]
    \centering
    \includegraphics[width=0.15\textwidth]{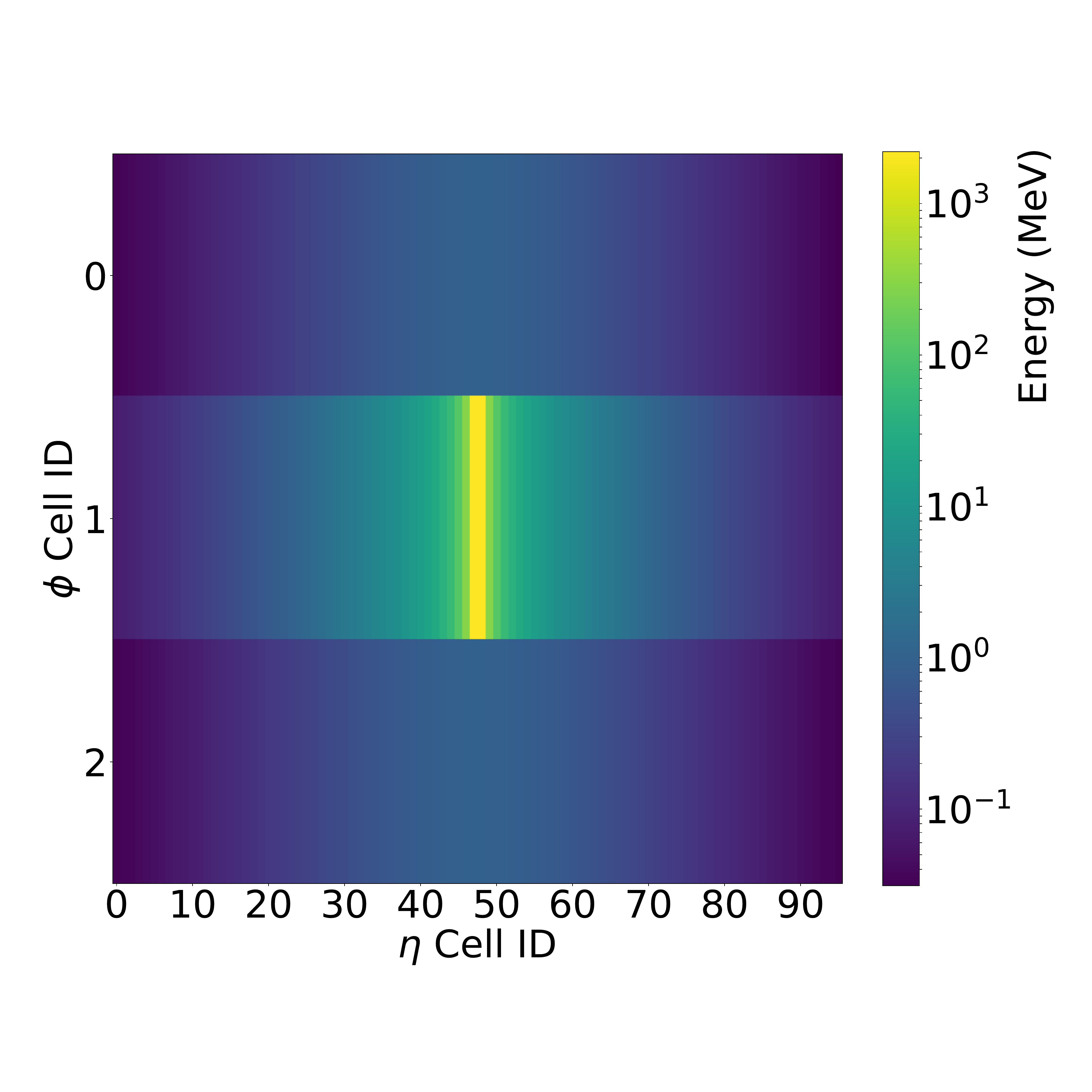}
    \includegraphics[width=0.15\textwidth]{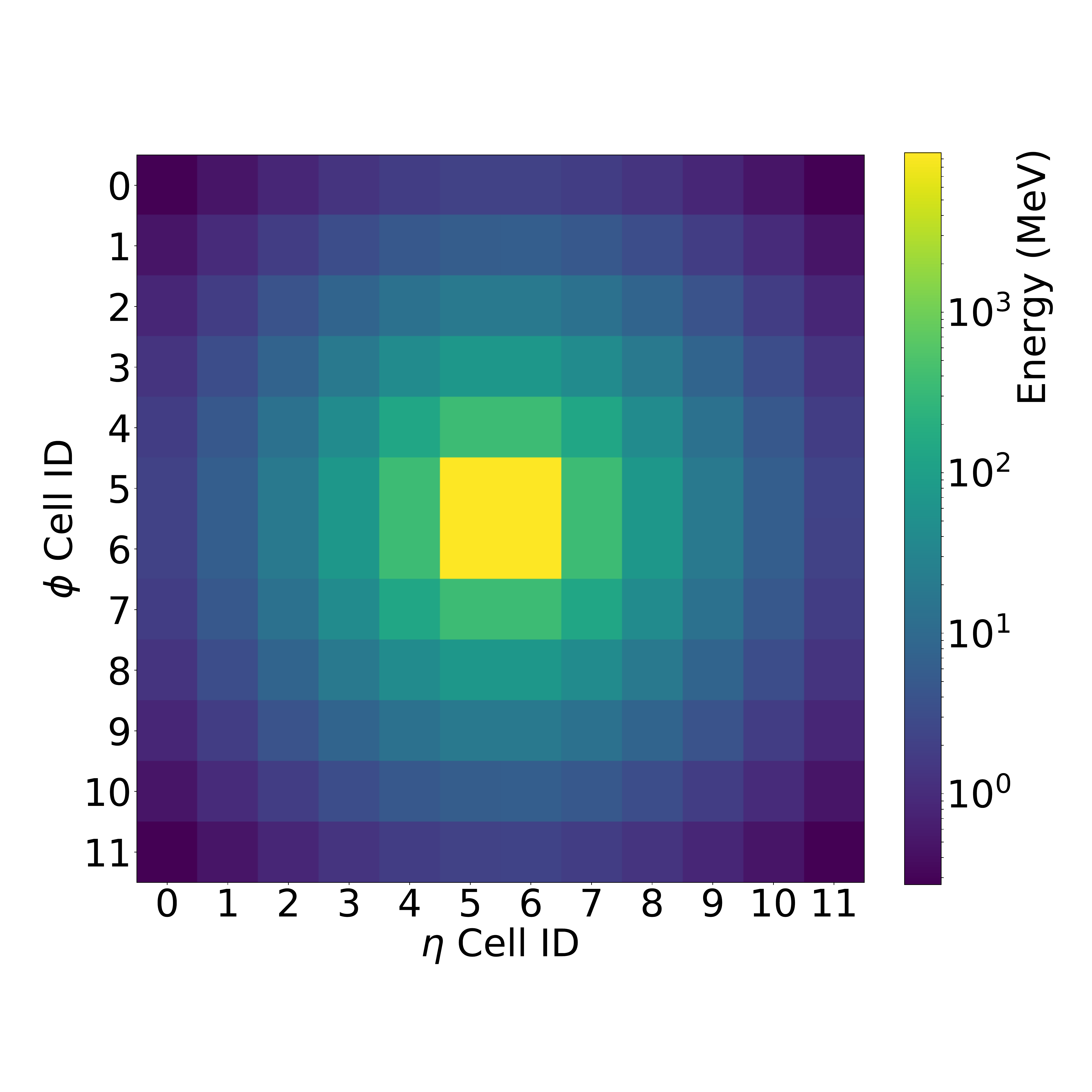}
    \includegraphics[width=0.15\textwidth]{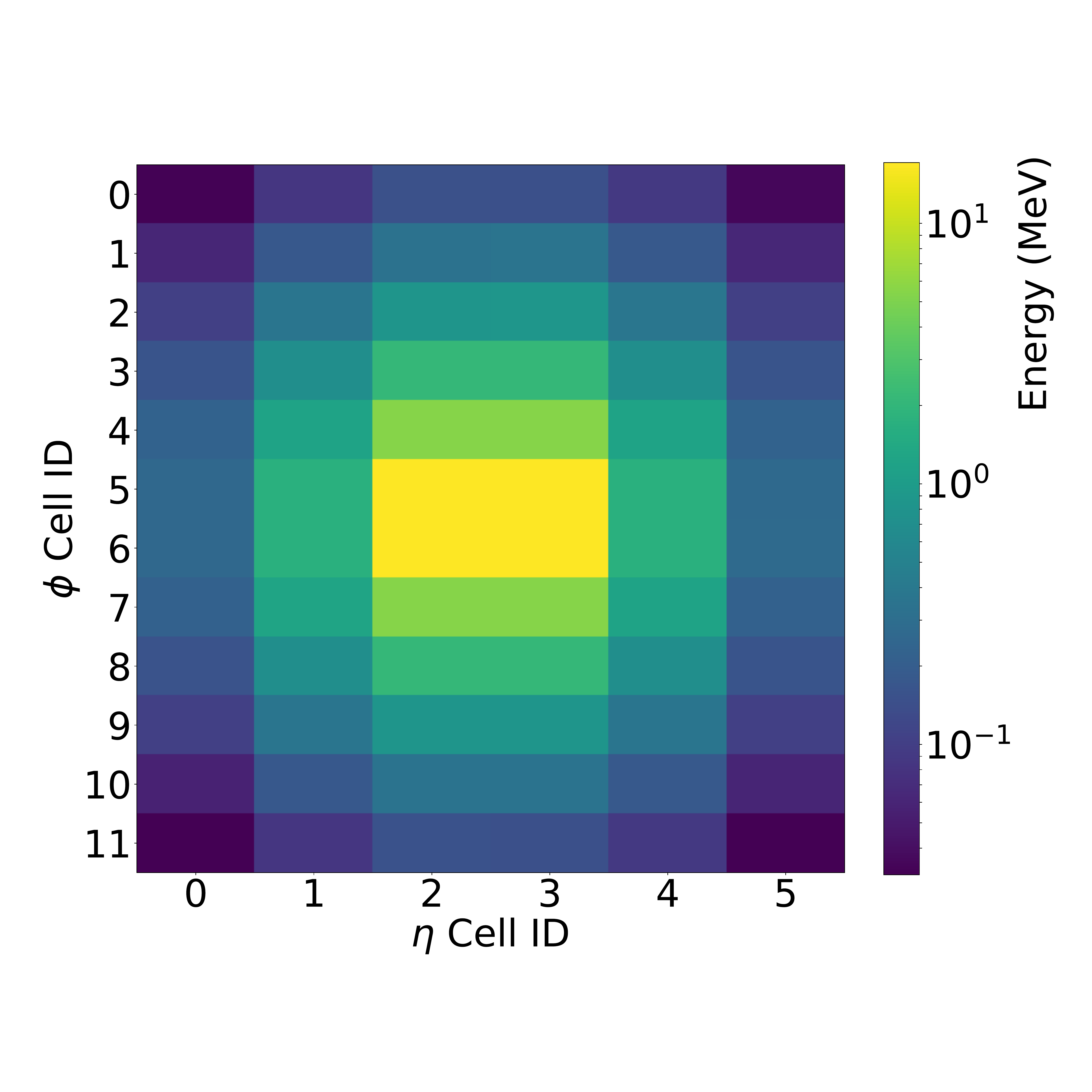}\\
    \includegraphics[width=0.15\textwidth]{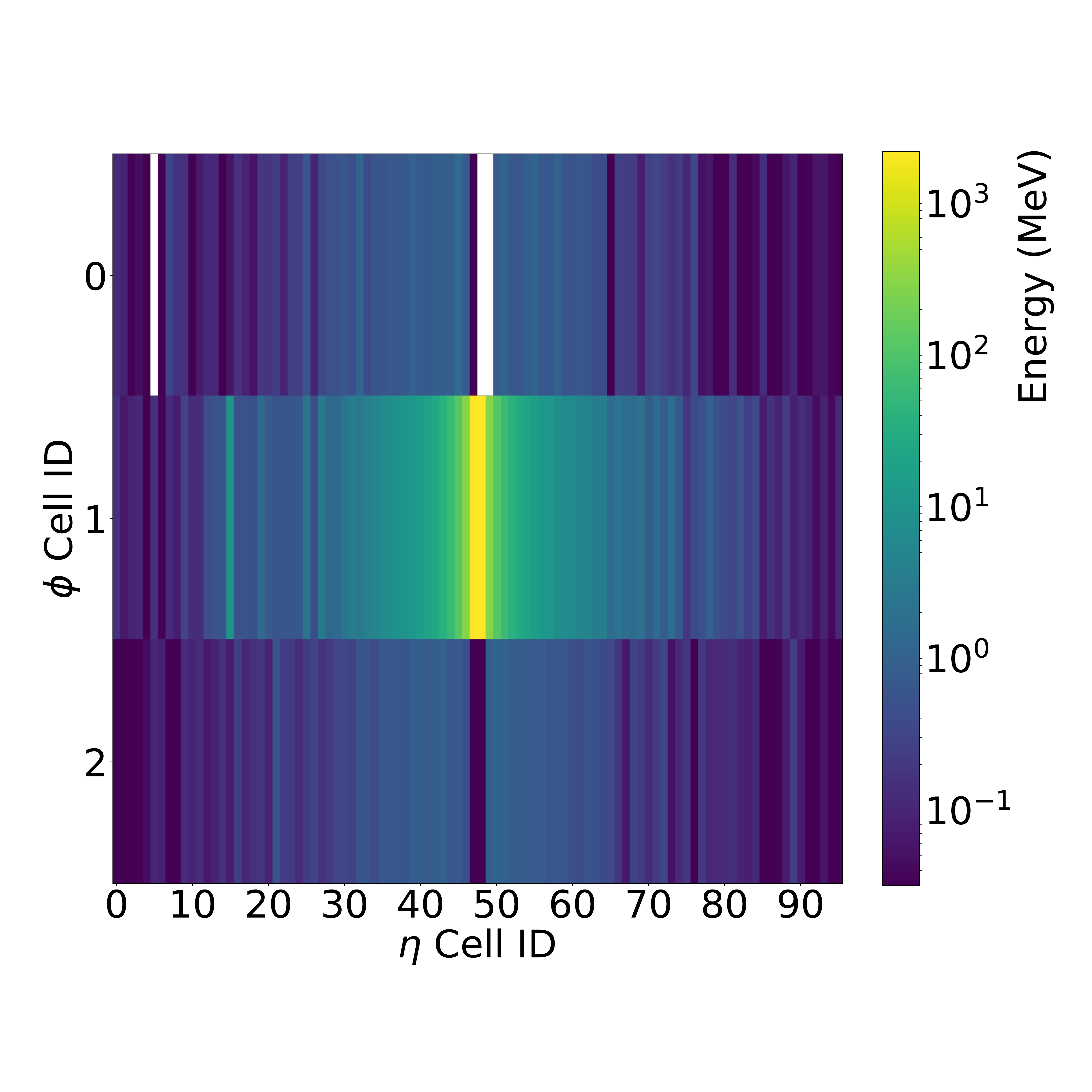}
    \includegraphics[width=0.15\textwidth]{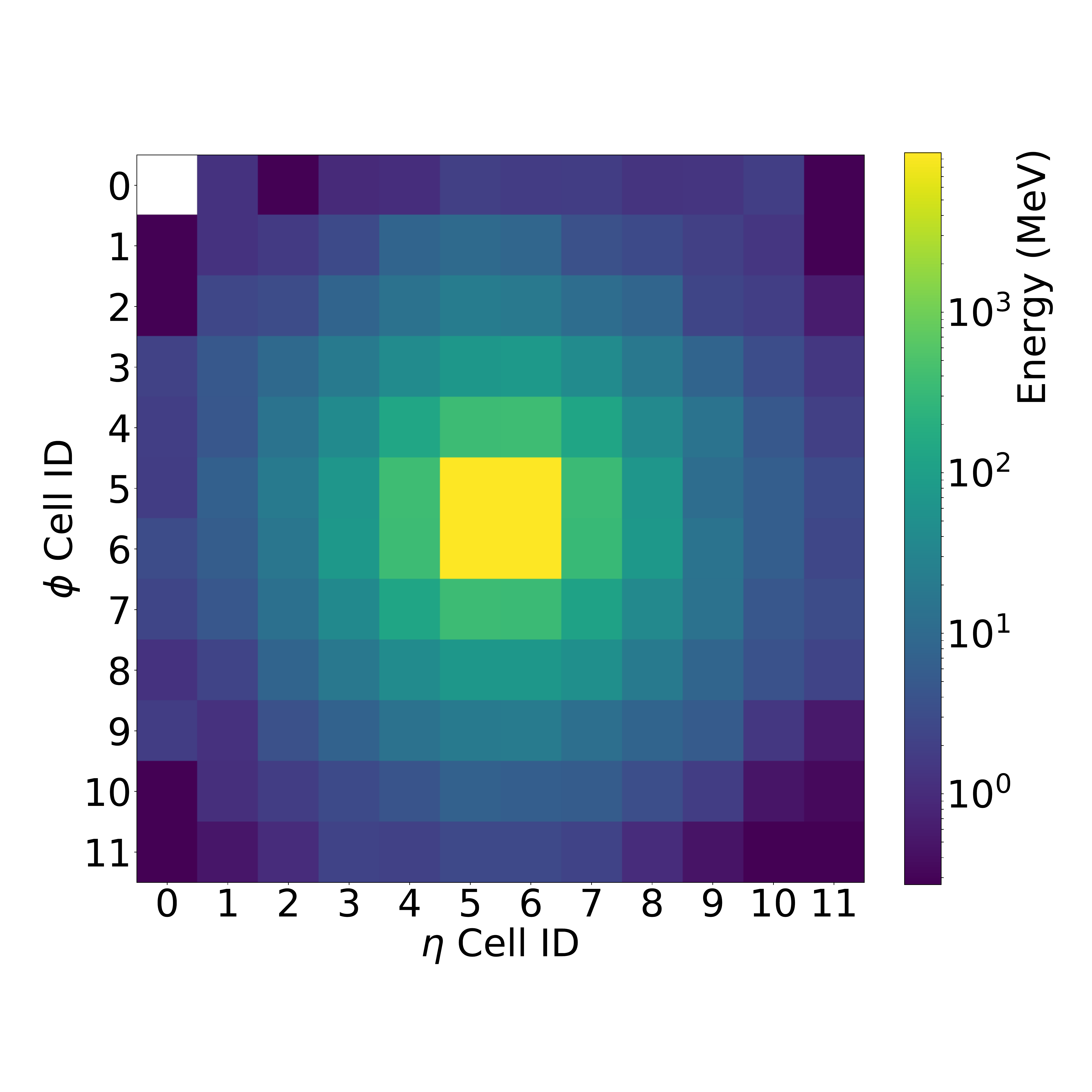}
    \includegraphics[width=0.15\textwidth]{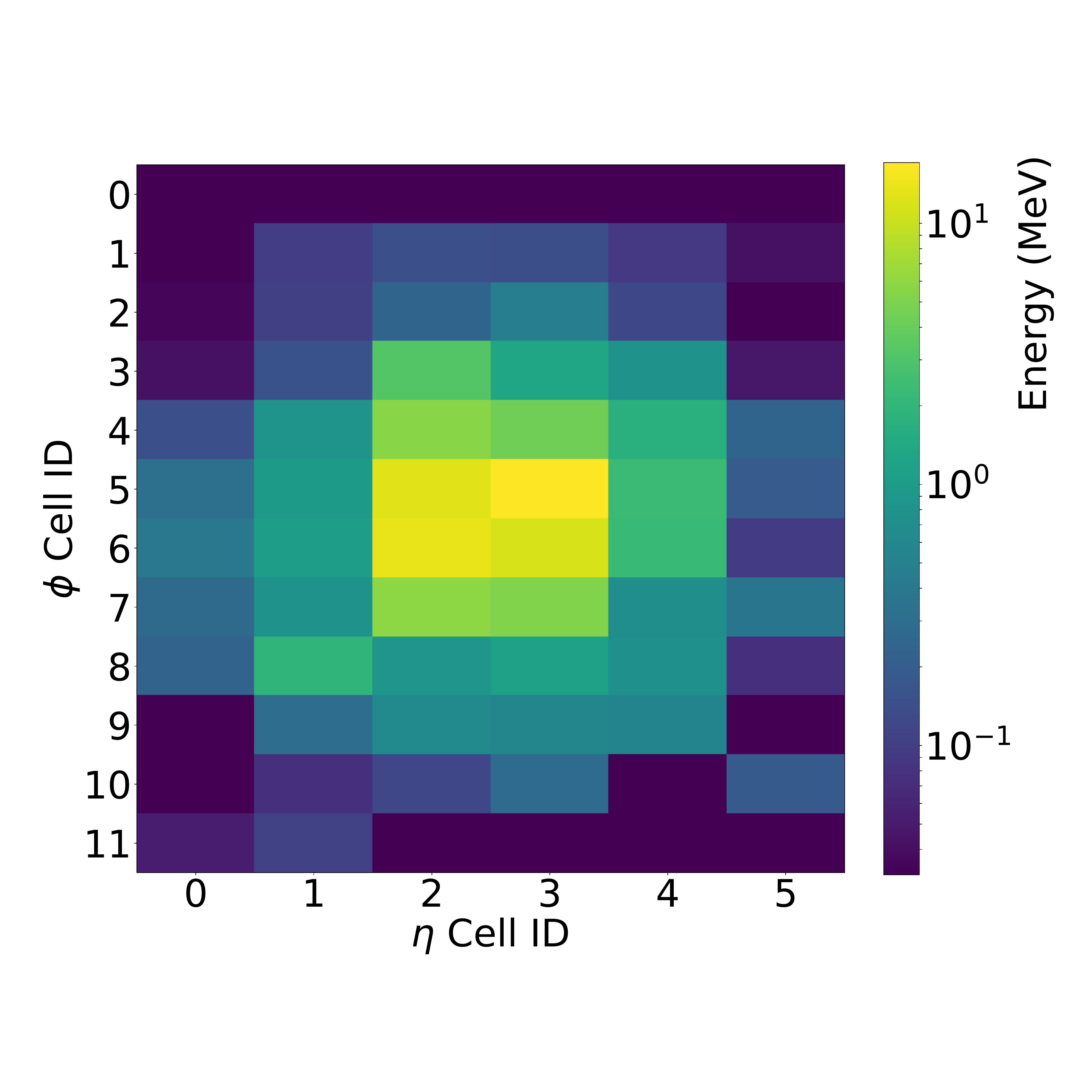}\\
    \caption{Average $\gamma$ \textsc{Geant4} shower (top), and average $\gamma$ \textsc{CaloGAN} shower (bottom), with progressive calorimeter depth (left to right).}
    \label{fig:gamma_avg}
\end{figure}

The average calorimeter deposition per voxel (Fig.~\ref{fig:gamma_avg}) suggests that the learned generative models of $\gamma$, $e^+$, and $\pi^+$ showers capture aspects of the underlying physical processes. For photon showers, for instance, the mean per-layer cell variations only show a $\sim 4\%$ and $\sim 1\%$ discrepancy in the first two layers where most energy is deposited for $e/\gamma$. This level of agreement is promising, but it is important to analyze more than the mean energy pattern to fully study the strengths and weaknesses of the proposed approach.


\begin{figure}[h]
    \centering
    \includegraphics[width=0.45\textwidth, trim={0cm, 0cm, 0cm, 0cm}]{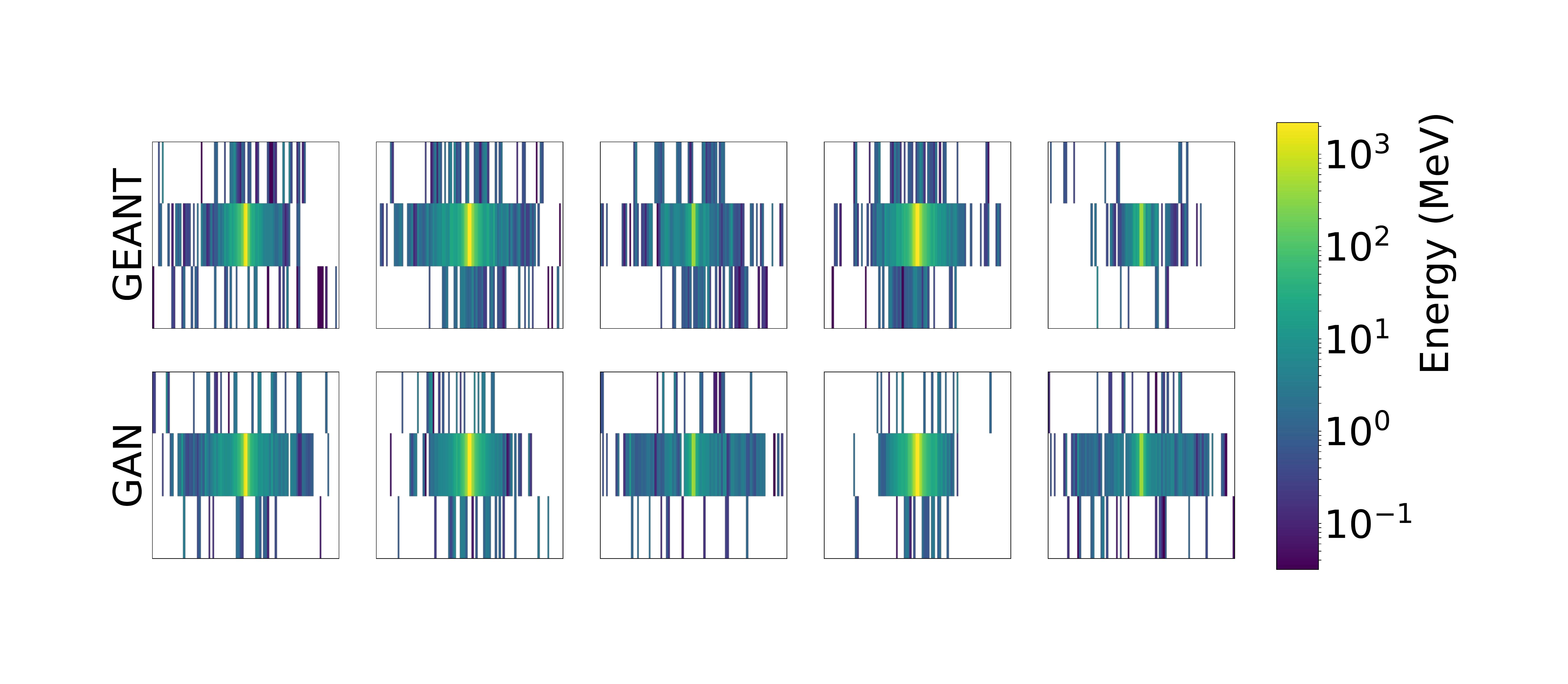}\hfill
    \includegraphics[width=0.45\textwidth, trim={0cm, 0cm, 0cm, 0cm}]{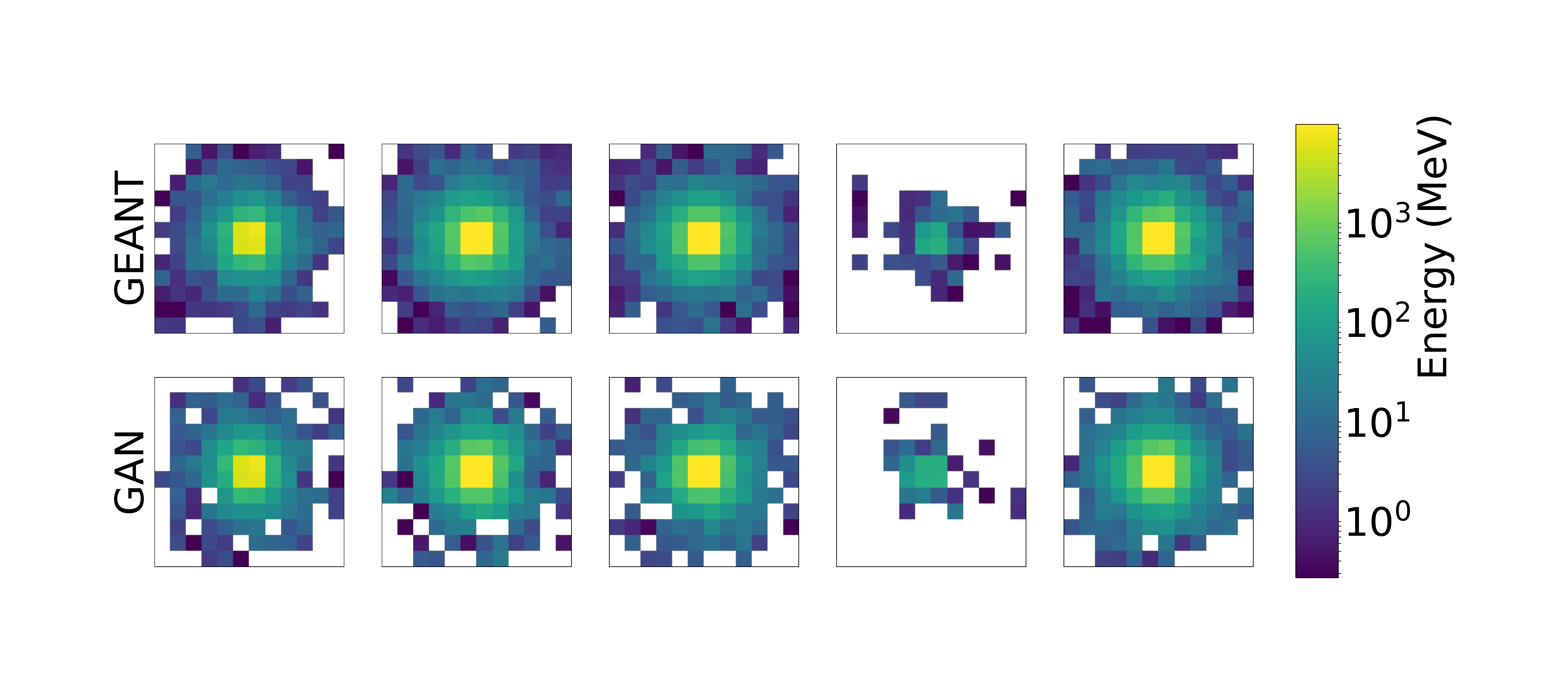}\hfill
    \includegraphics[width=0.45\textwidth, trim={0cm, 0cm, 0cm, 0cm}]{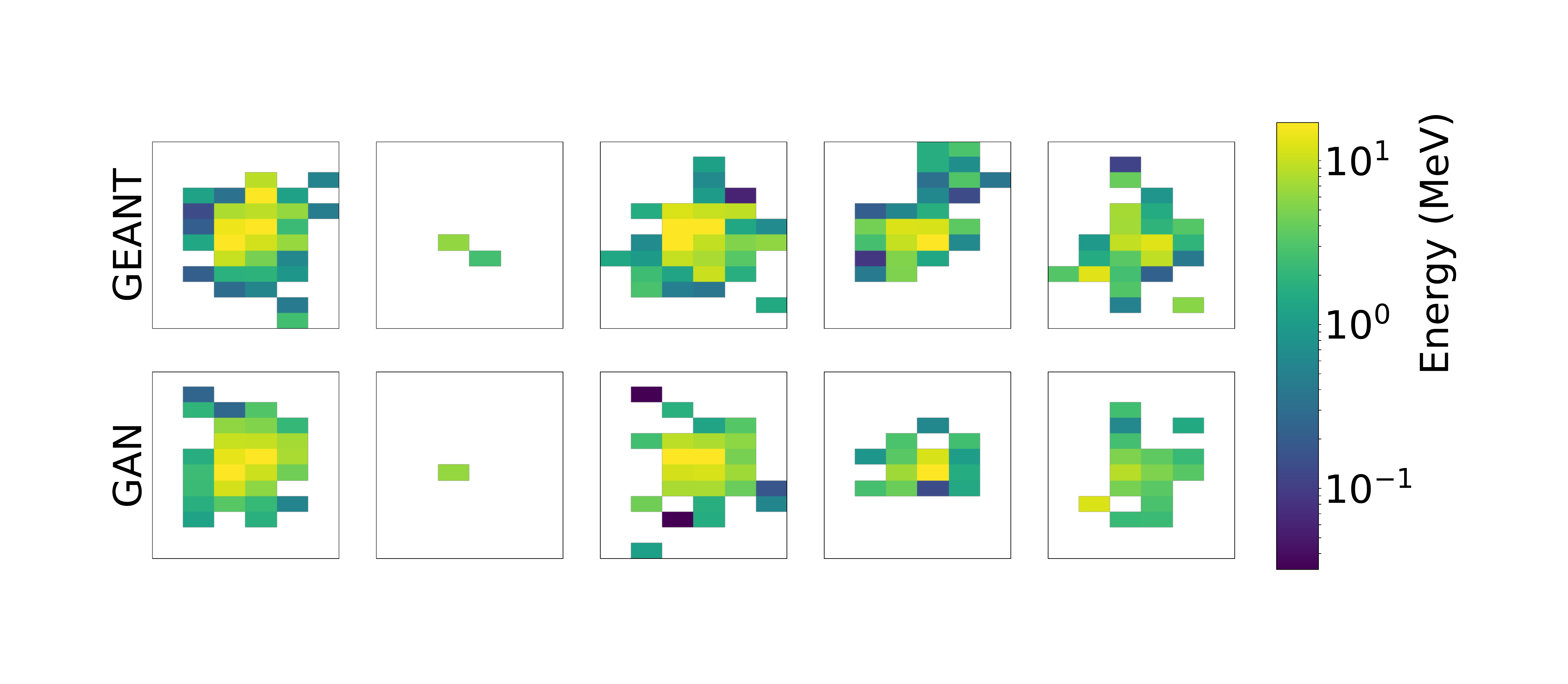}\\
    \caption{Five randomly selected $\gamma$ showers per calorimeter layer from \textsc{Geant4} (top) and their five nearest neighbors (by euclidean distance) from a set of \textsc{CaloGAN} candidates.}
    \label{fig:gamma_nn}
\end{figure}

The \textsc{CaloGAN}-generated samples are checked for adequate diversity and lack of direct memorization of the \textsc{Geant4} samples used for training. The nearest (by Euclidean distance) \textsc{Geant4} image is found for each of a random selection of \textsc{CaloGAN} images in order to verify the desired characteristics (Fig.~\ref{fig:gamma_nn}). The samples show strong inter- and intra-class diversity and no evidence of memorization since the closest images do not look exactly the same.

\subsection{Shower Shape Description}

\begin{figure*}[!ht]
    \centering

    \includegraphics[width=0.2\textwidth]{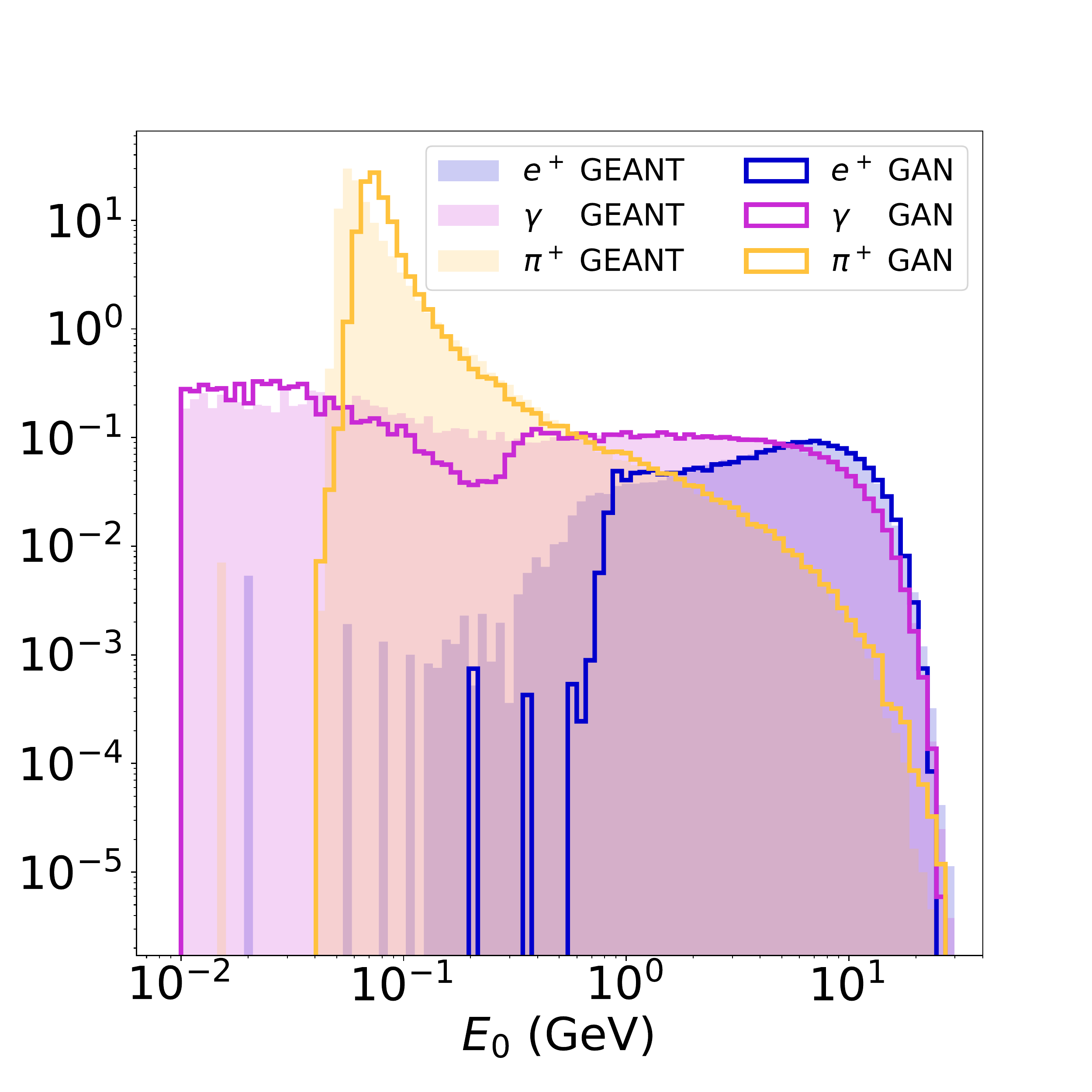}
    \includegraphics[width=0.2\textwidth]{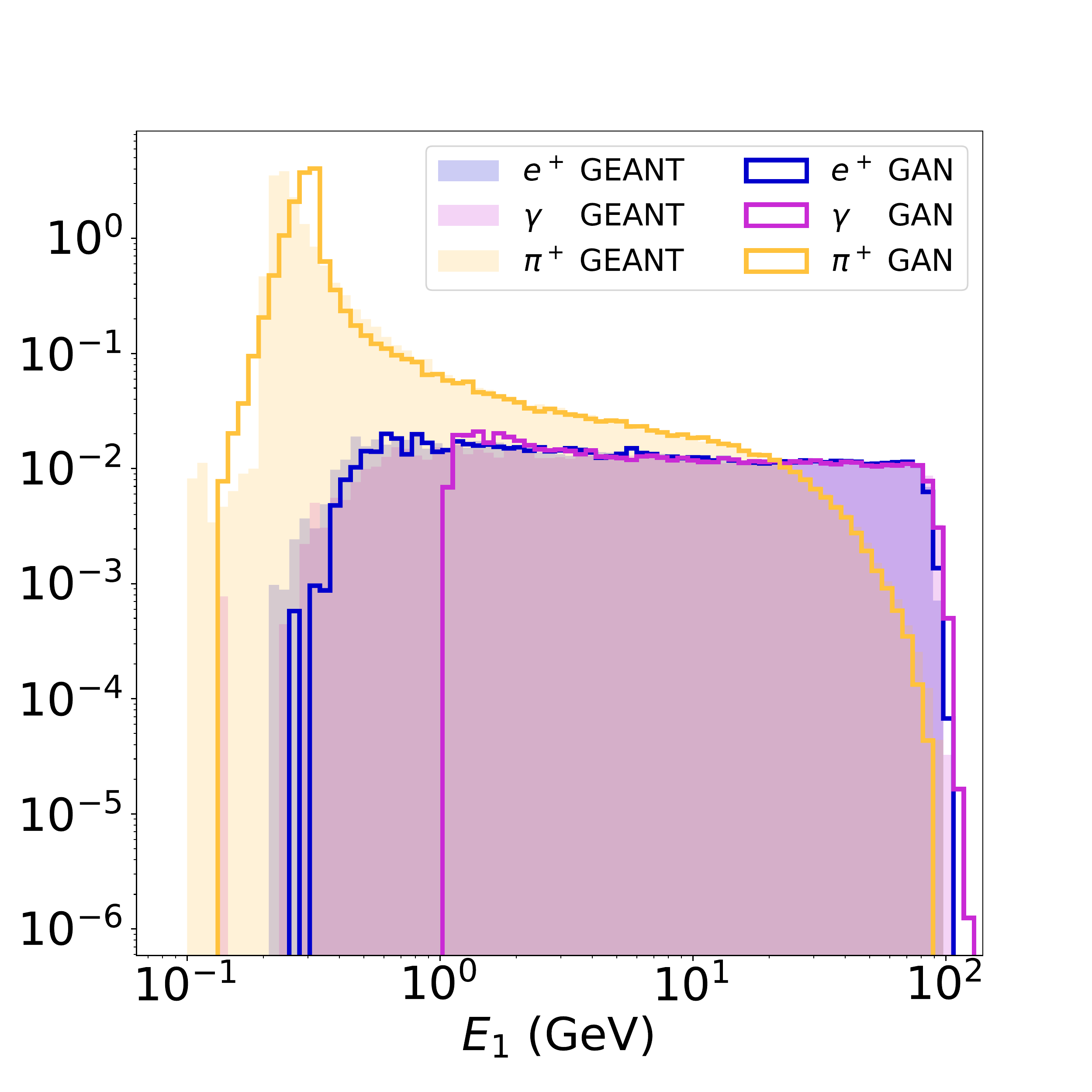}
    \includegraphics[width=0.2\textwidth]{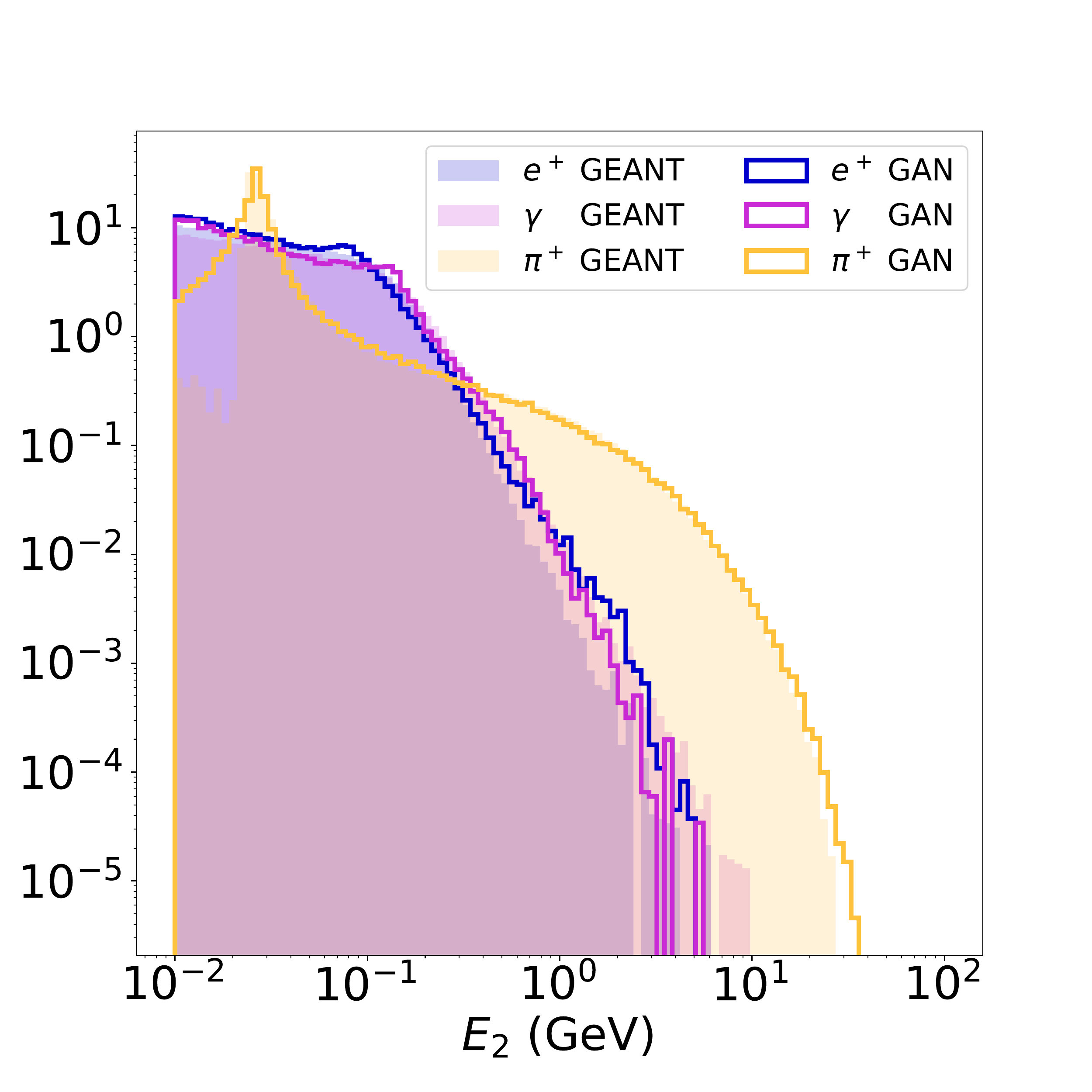}
    \includegraphics[width=0.2\textwidth]{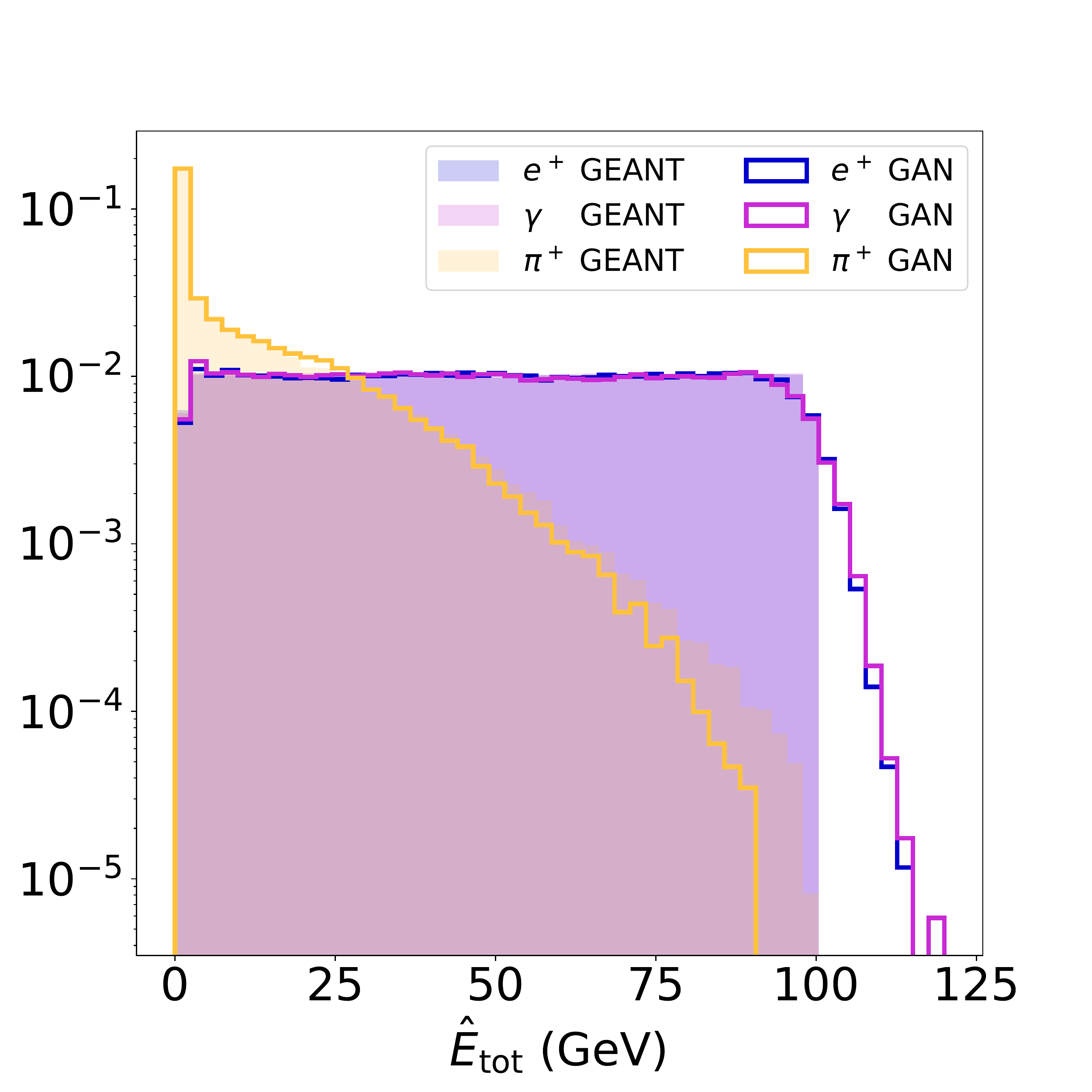}
    
    \includegraphics[width=0.2\textwidth]{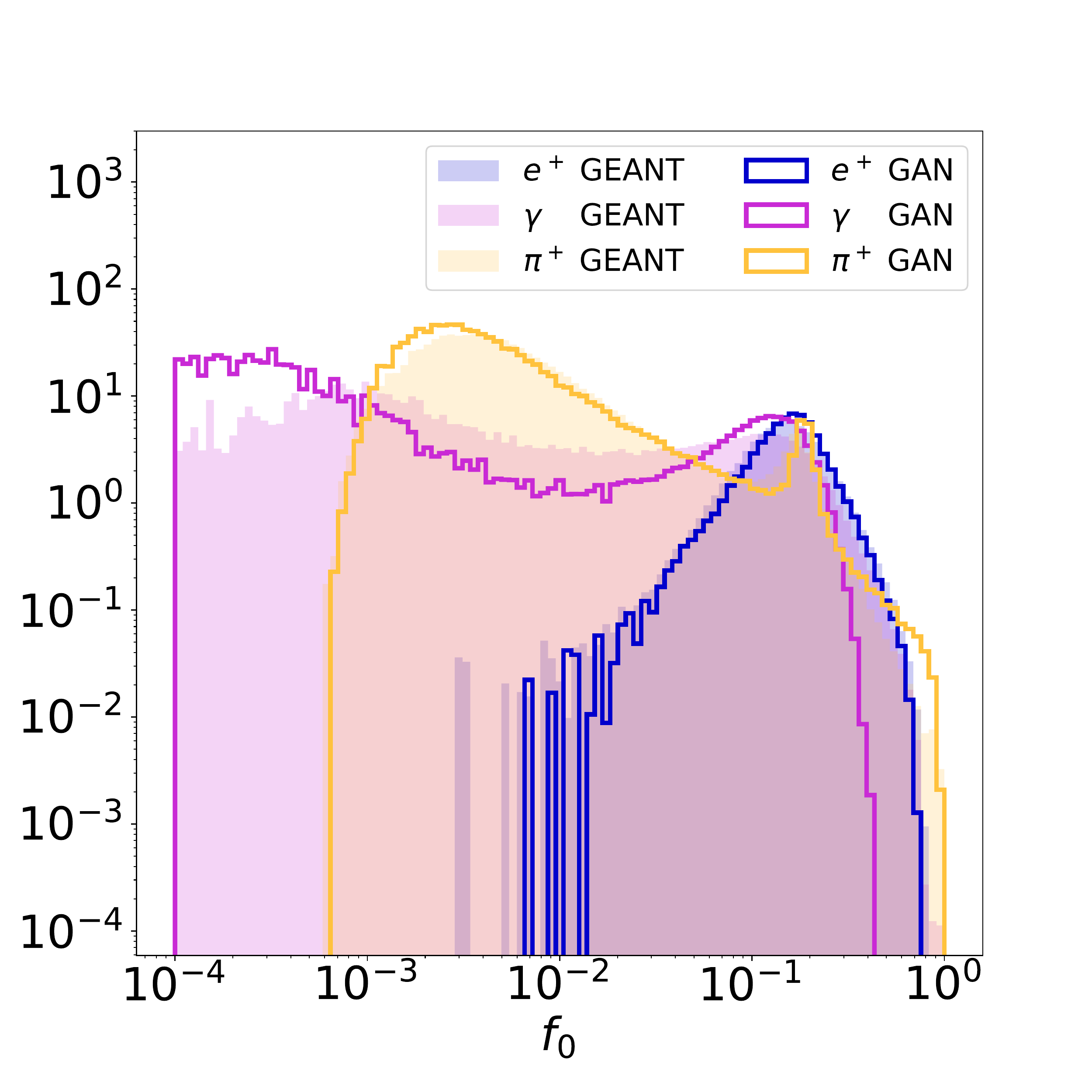}
    \includegraphics[width=0.2\textwidth]{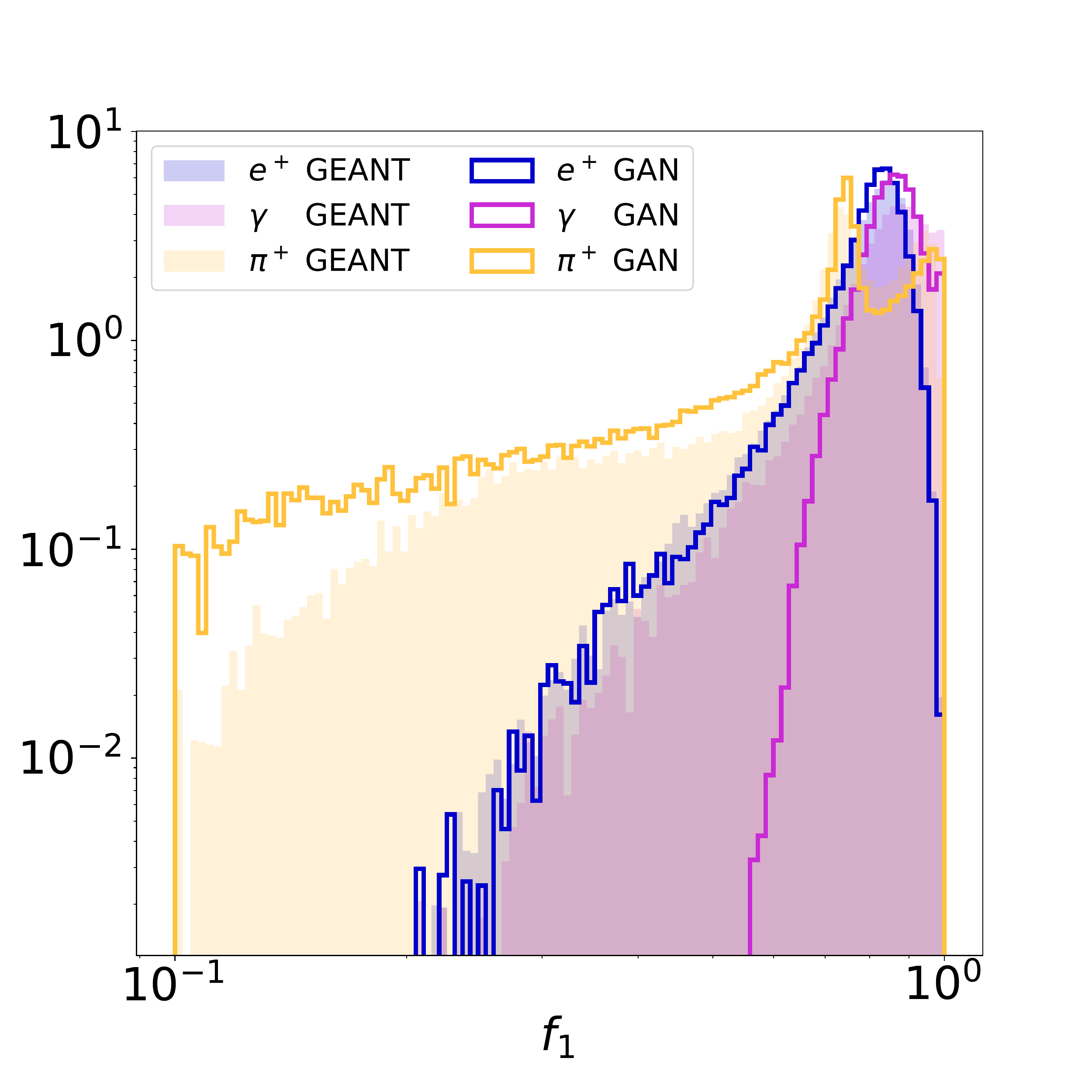}
    \includegraphics[width=0.2\textwidth]{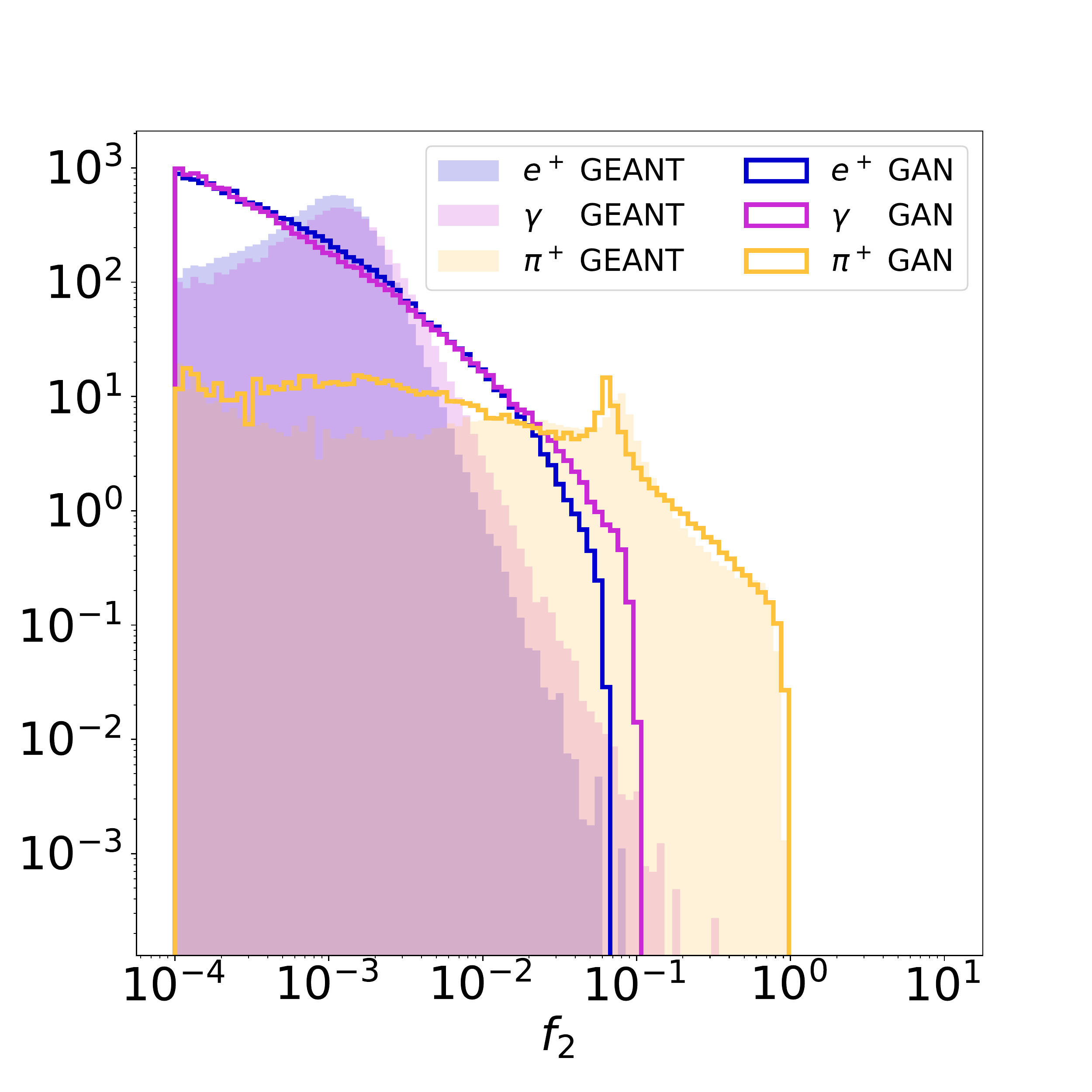}
    \includegraphics[width=0.2\textwidth]{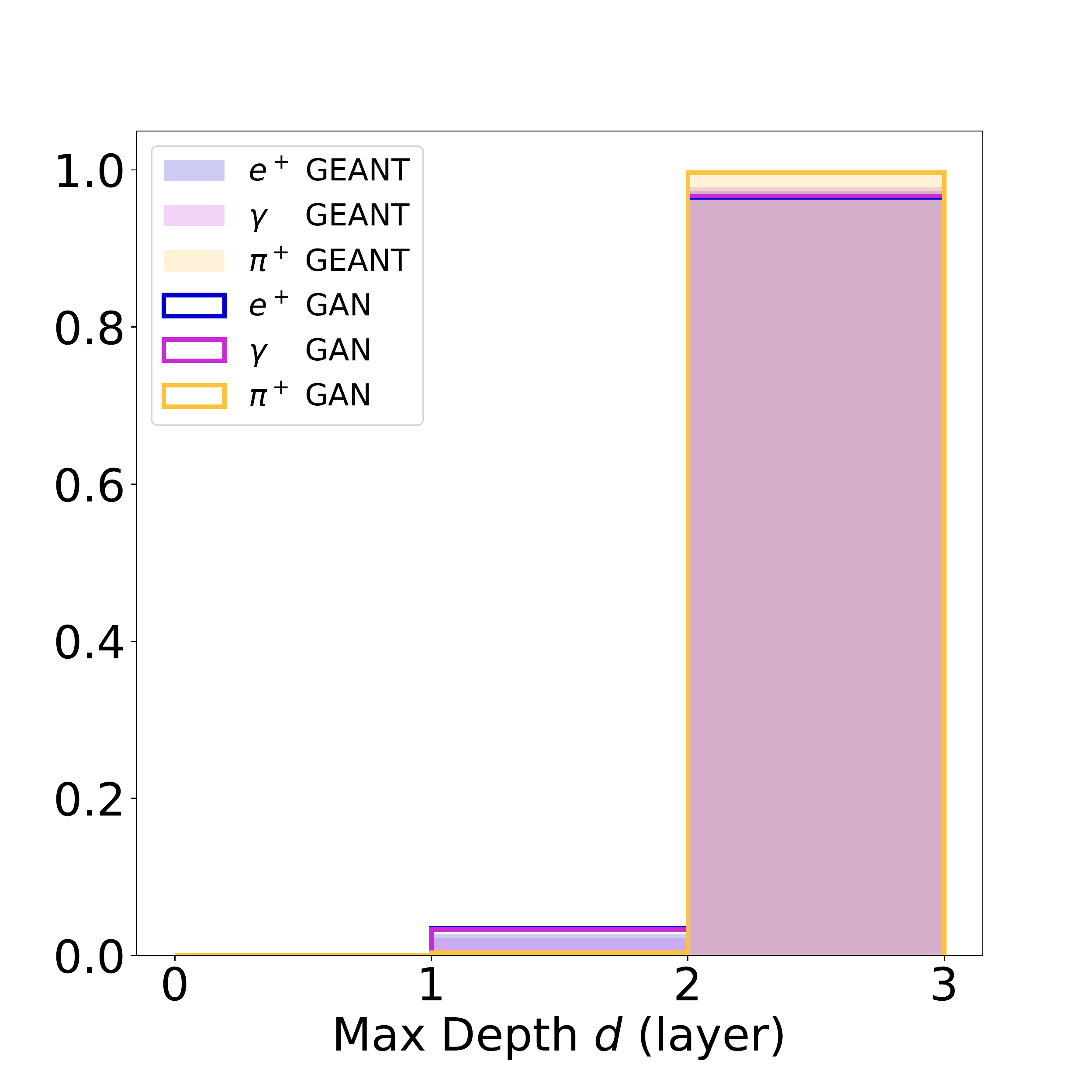}
    
    \includegraphics[width=0.2\textwidth]{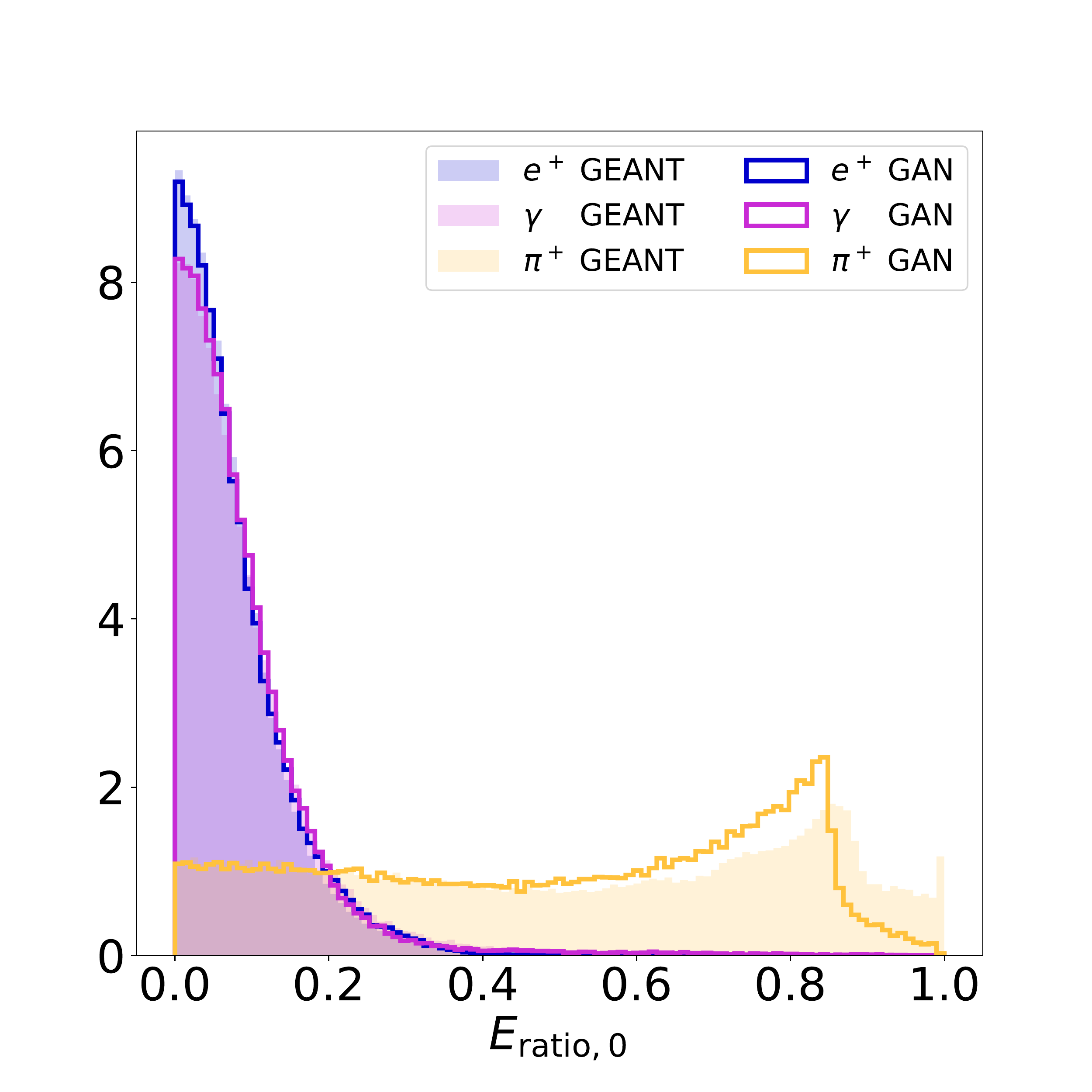}
    \includegraphics[width=0.2\textwidth]{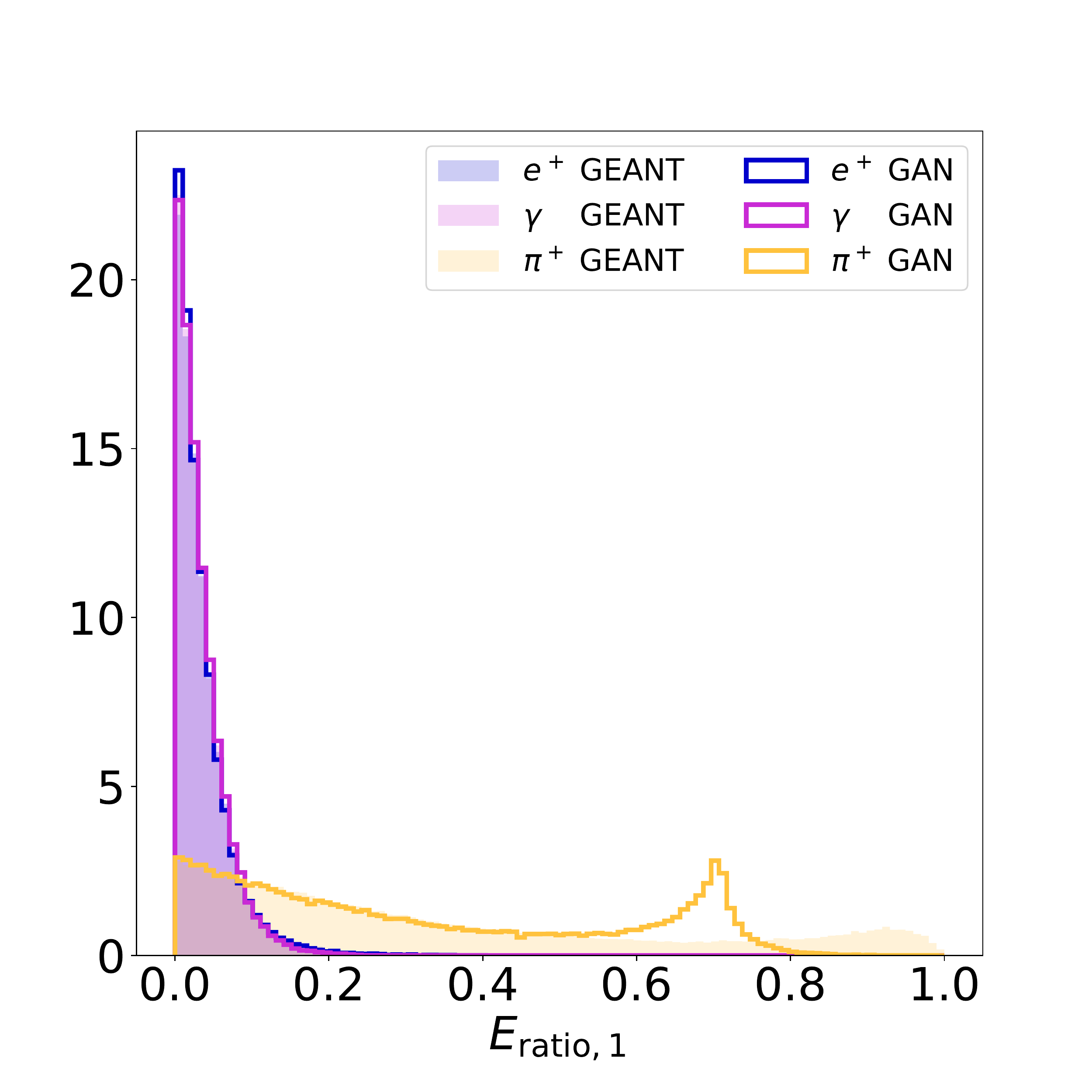}
    \includegraphics[width=0.2\textwidth]{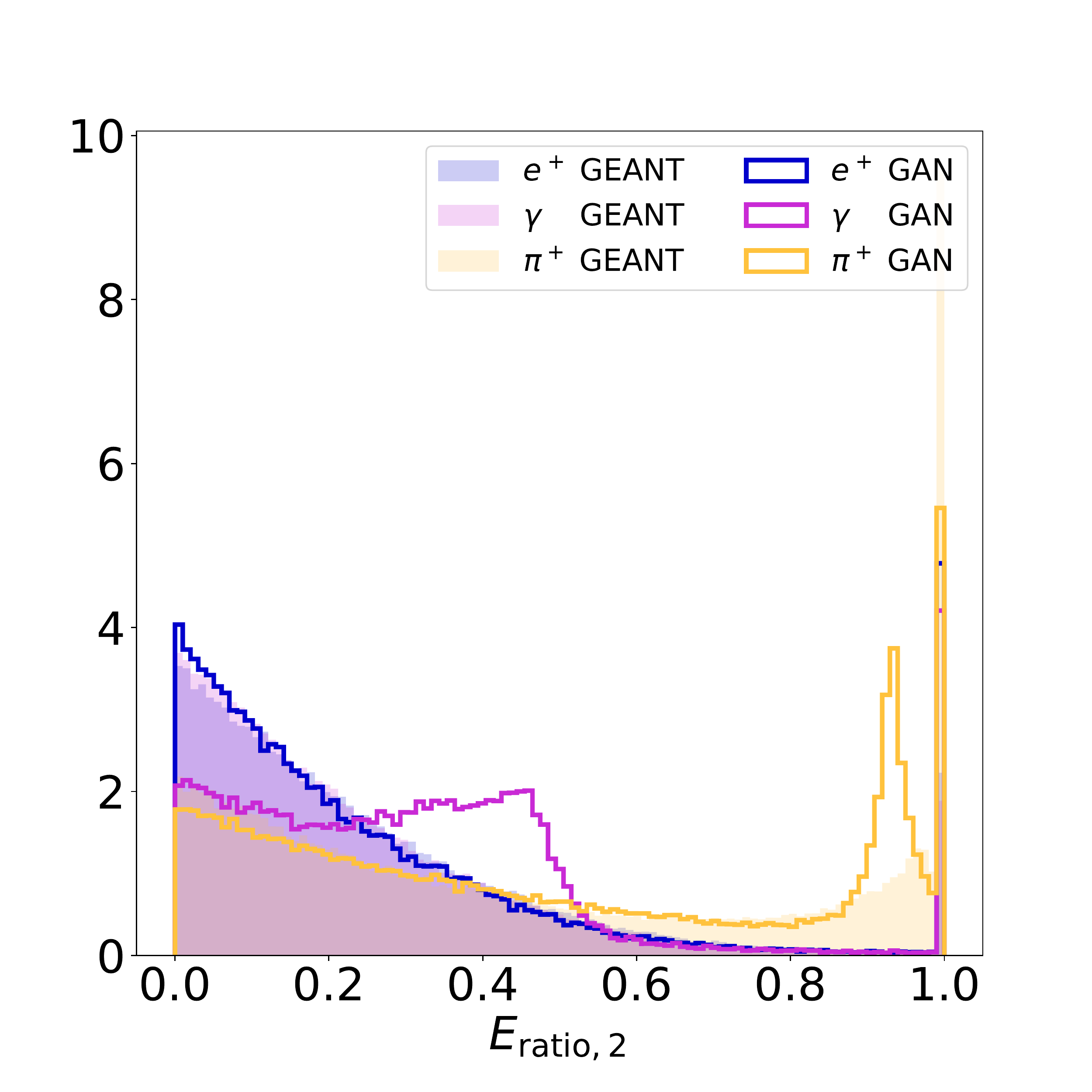}
    \includegraphics[width=0.2\textwidth]{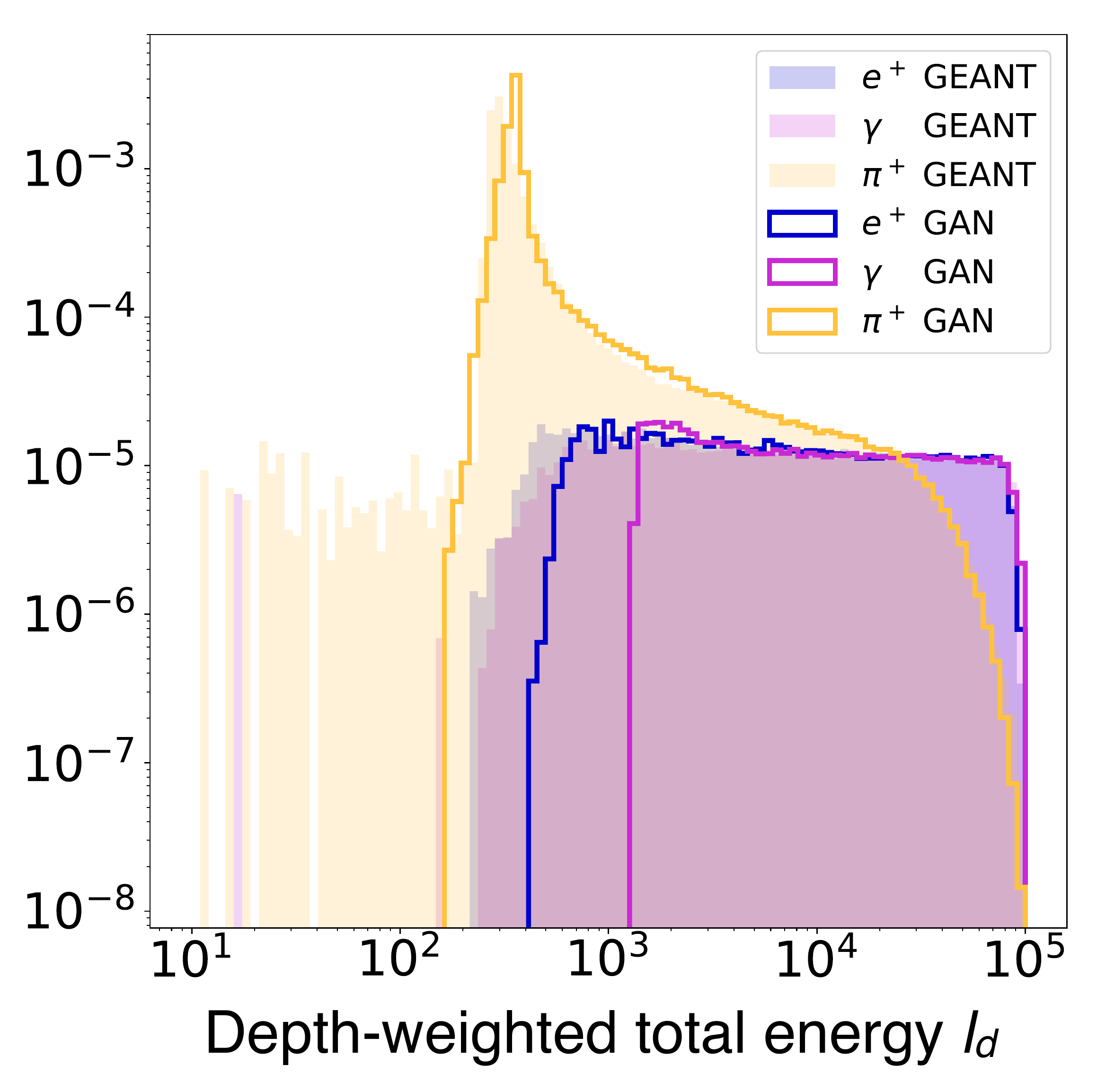}
    
    \includegraphics[width=0.2\textwidth]{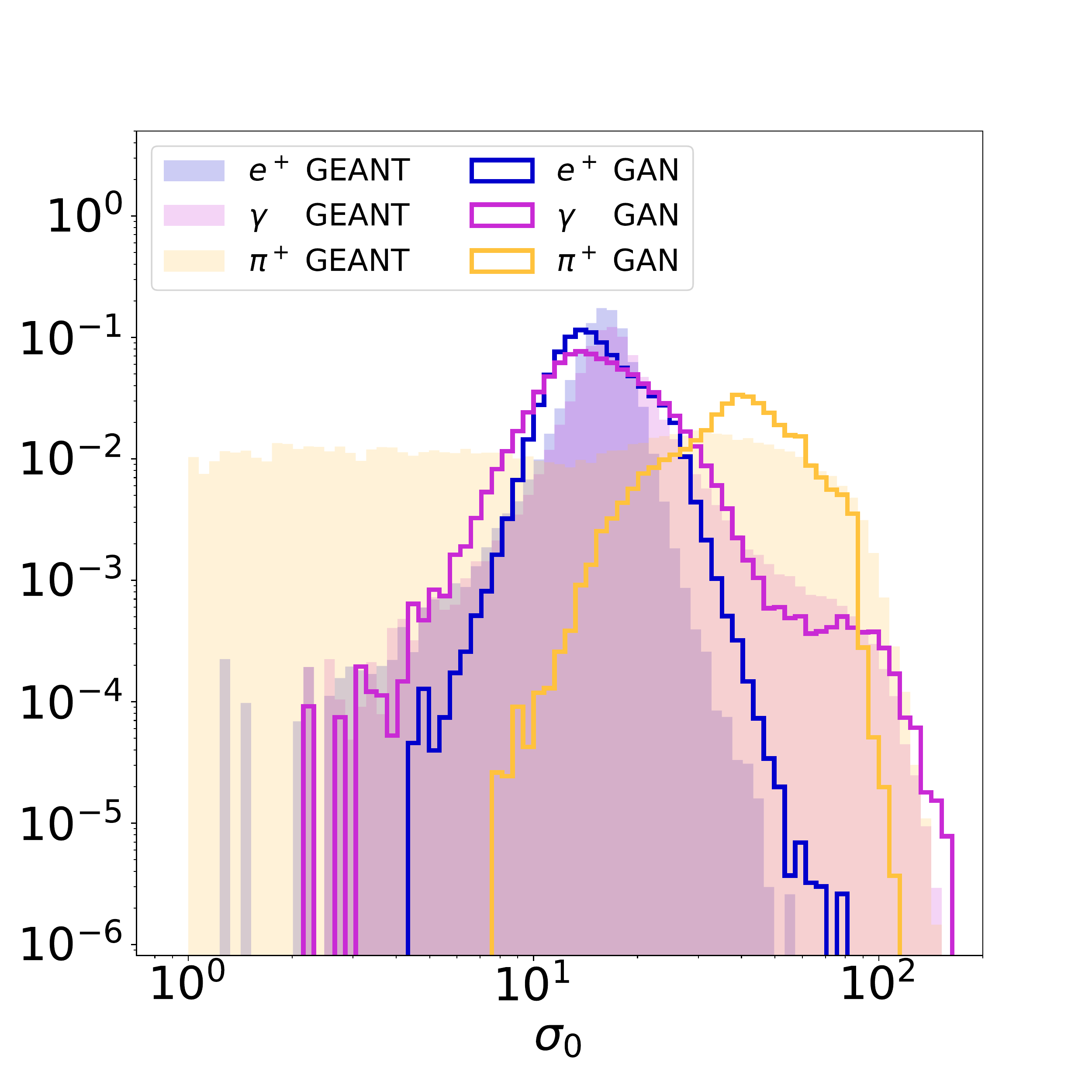}
    \includegraphics[width=0.2\textwidth]{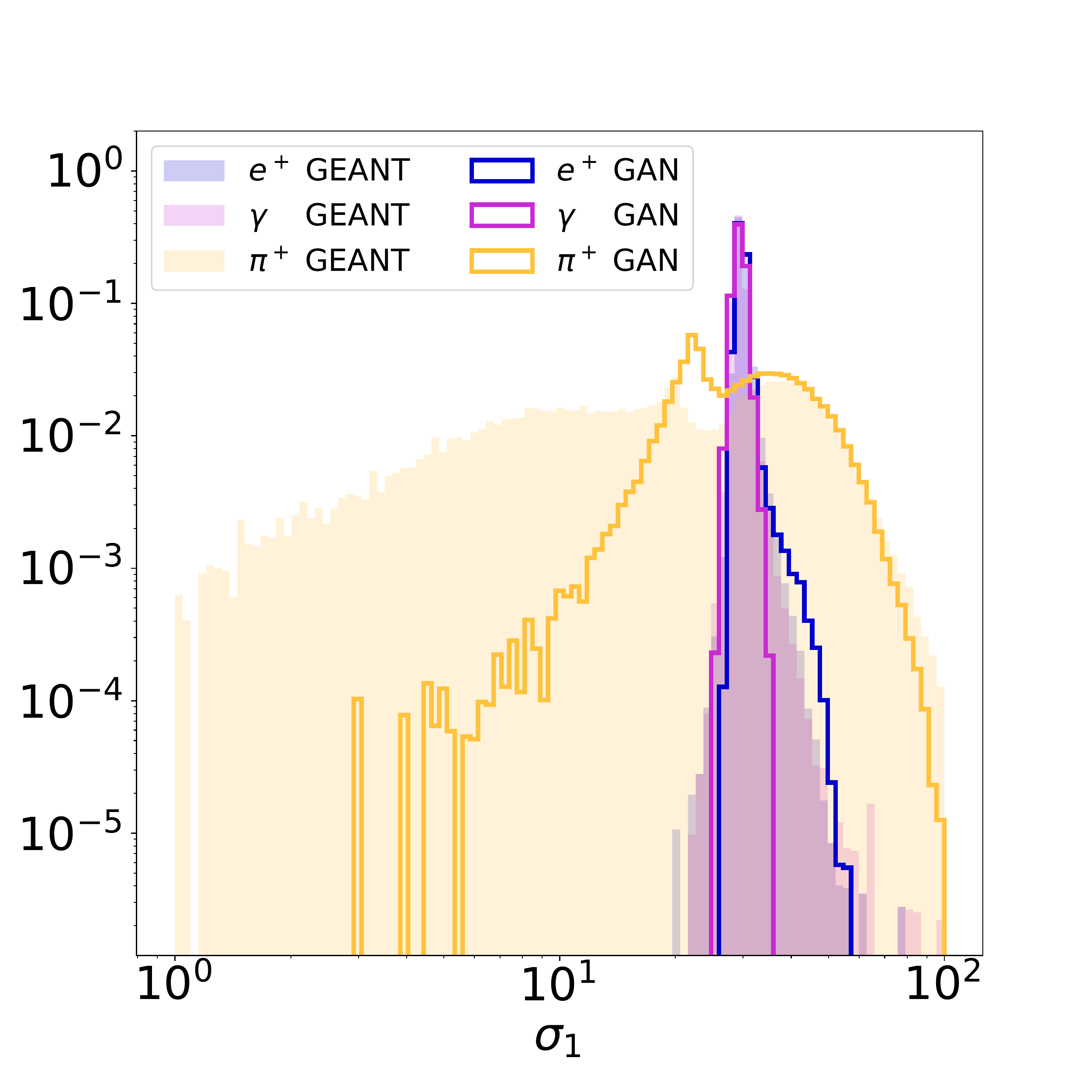}
    \includegraphics[width=0.2\textwidth]{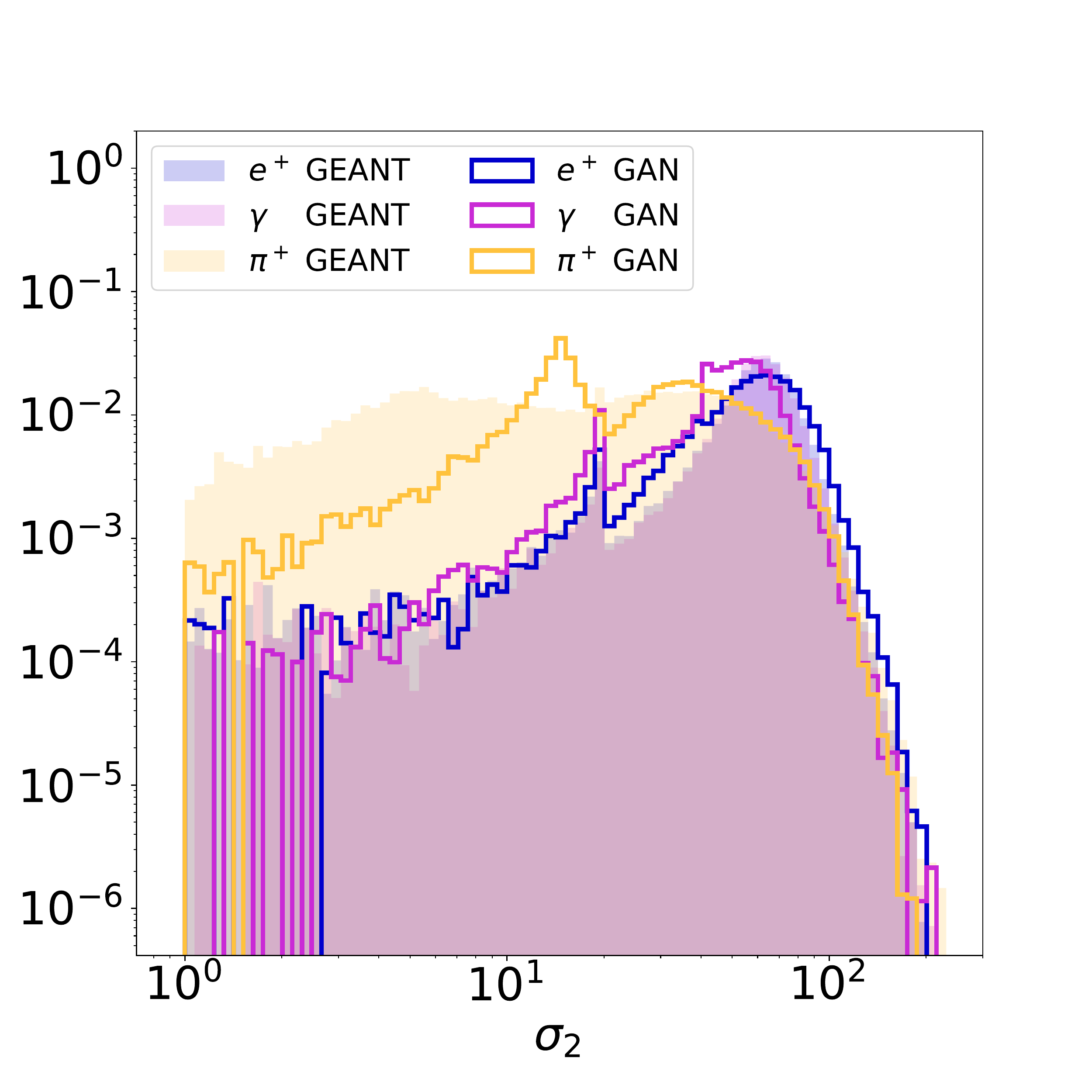}
    \includegraphics[width=0.2\textwidth]{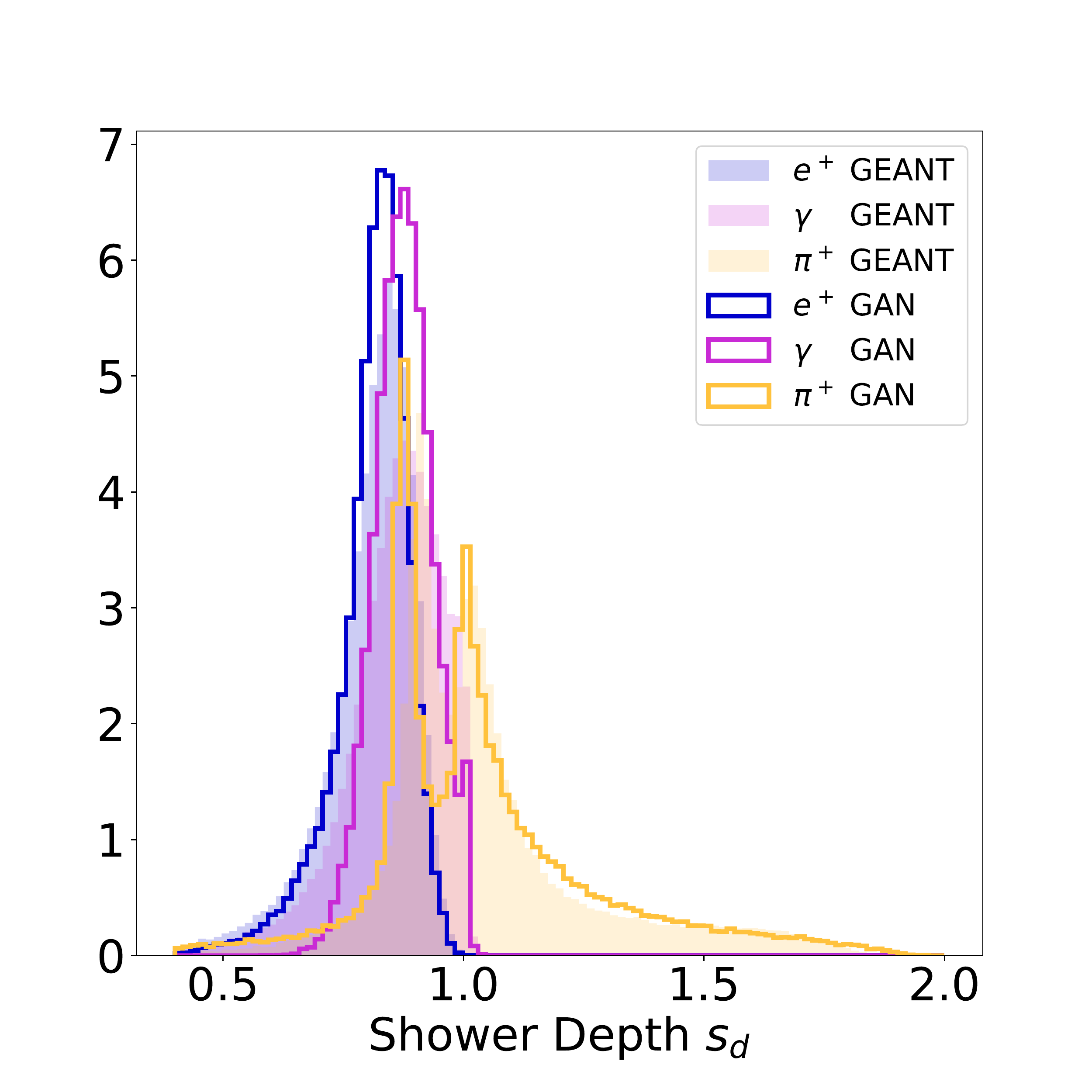}
    
    \includegraphics[width=0.2\textwidth]{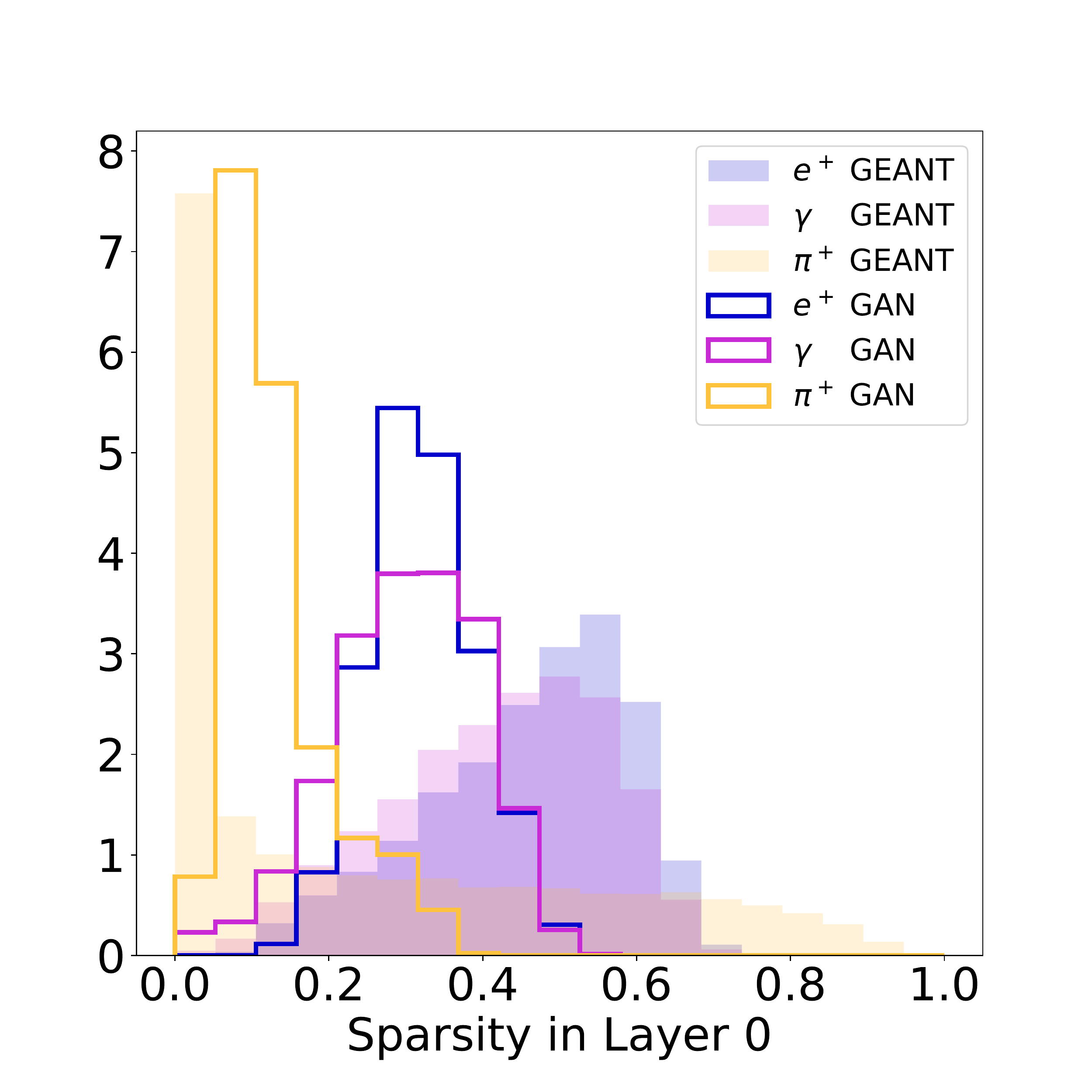}
    \includegraphics[width=0.2\textwidth]{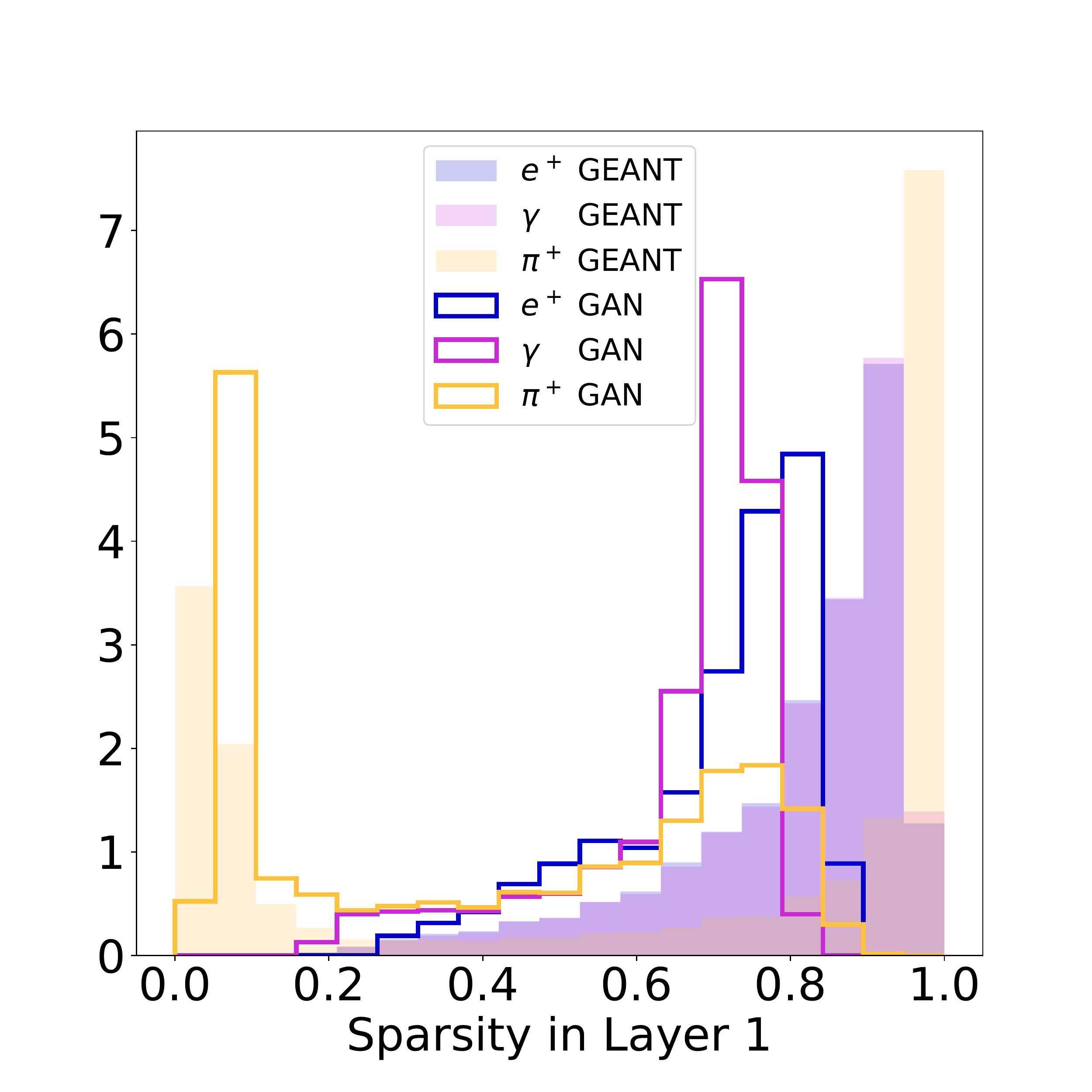}
    \includegraphics[width=0.2\textwidth]{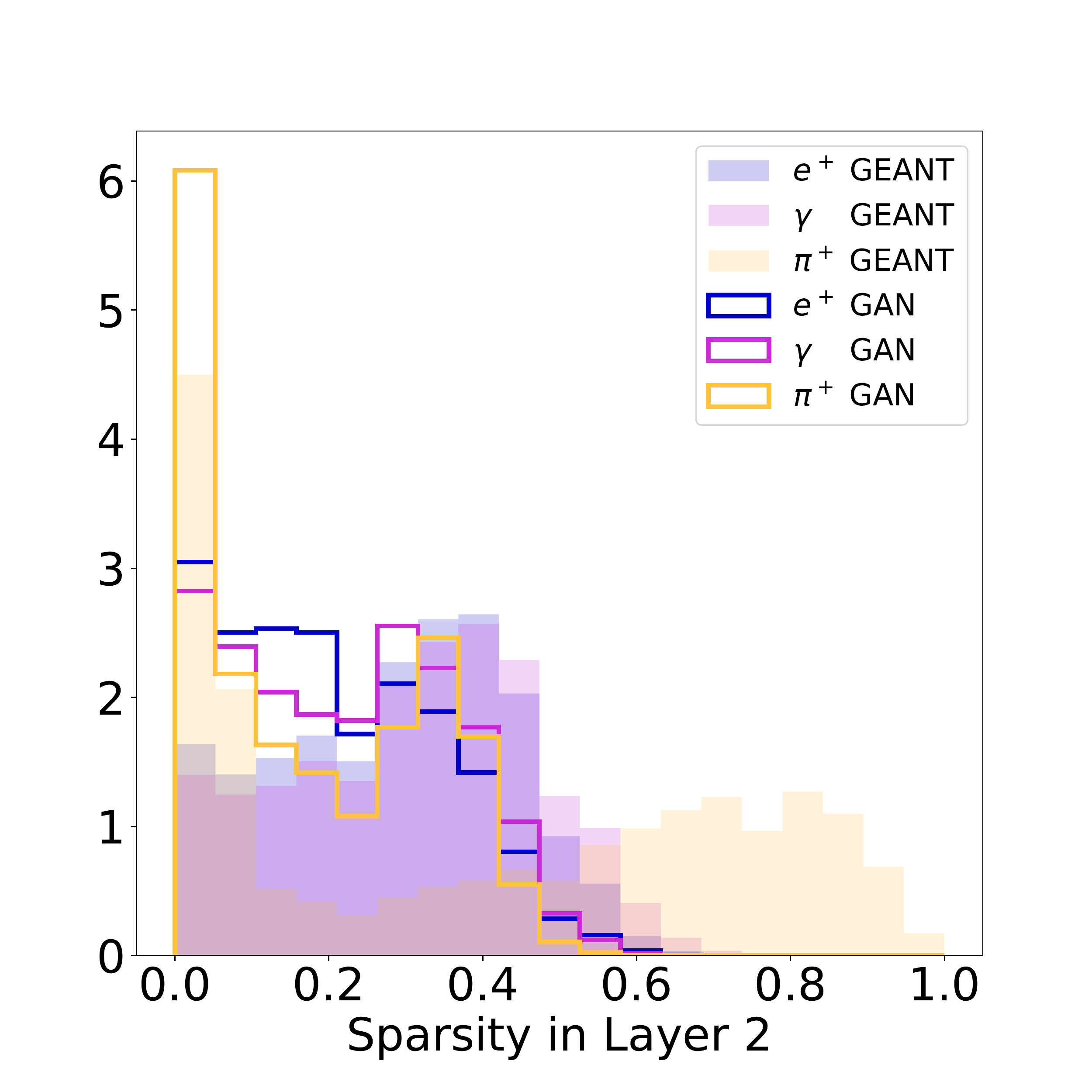}
    \includegraphics[width=0.2\textwidth]{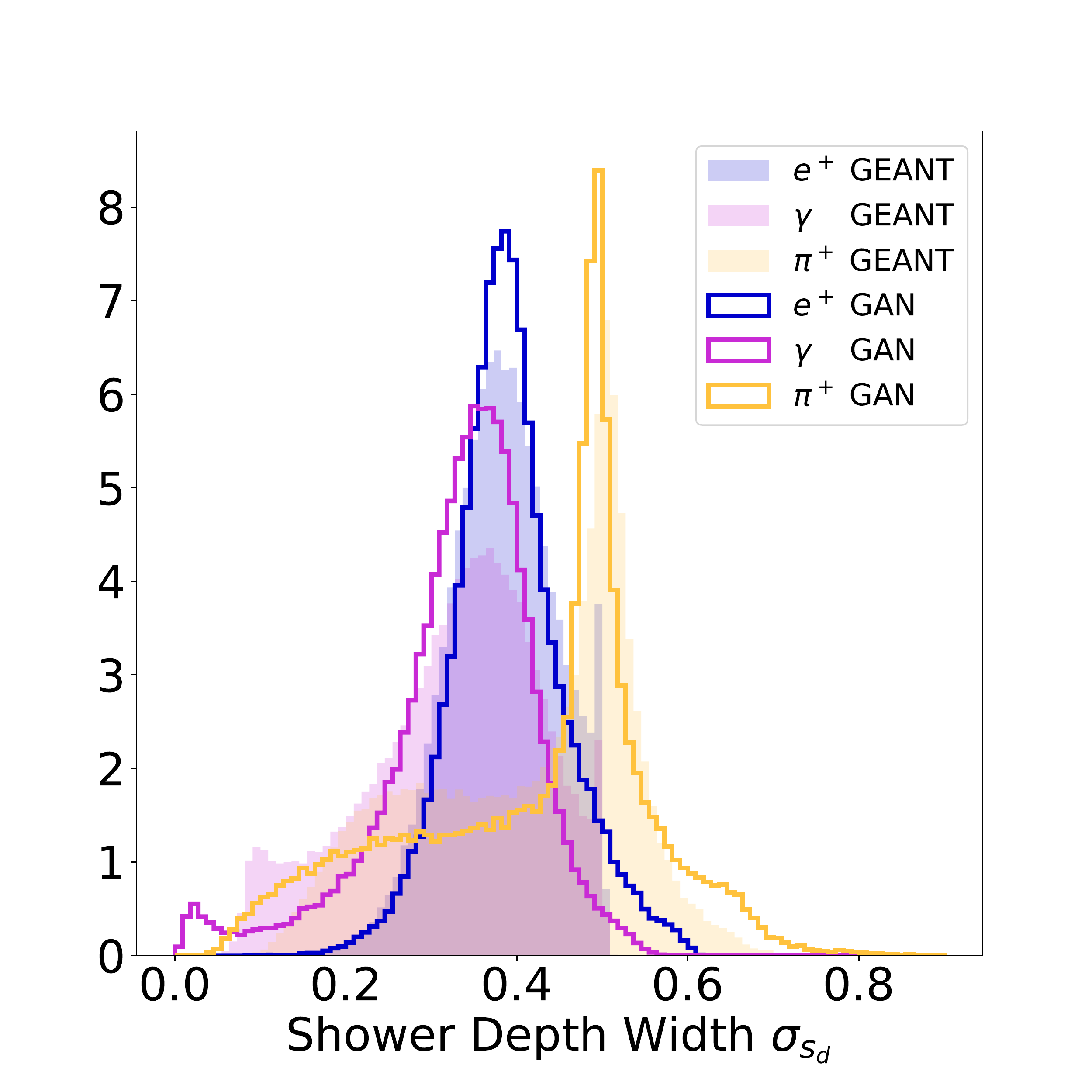}
    \caption{Comparison of shower shape variables and other variables of interest, such as the sparsity level per layer, for the \textsc{Geant4} and \textsc{CaloGAN} datasets for $e^+$, $\gamma$ and $\pi^+$. See [PRD companion paper] for detailed definitions.}  
    \label{fig:shower_shapes}
\end{figure*}

Geometrically and physically motivated shower shape variables~\cite{Agashe:2014kda} are used as further validation and introspection into the capabilities of the \textsc{CaloGAN} to adequately model and capture non-linear functional representations of the simulated data distribution (Fig.~\ref{fig:shower_shapes}). In fact, it is desirable for the \textsc{CaloGAN} to recover the target distribution of these 1D statistics. 

The network is not shown any shower shape variable (only pixel values) at training time - therefore, it is encouraging to note that the \textsc{CaloGAN} recovers the simulated data distribution for a variety of shower shapes across the three particle types. However, certain features of some distributions are not well-described. This is a challenge for the future and will likely require improvements to the architecture and training procedure. Longer trainings of higher capacity architectures have shown promise in rectifying some of these issues.

Examining 1D statistics does not probe correlations between shower shapes or higher dimensional aspects of the probability distribution.  One way to examine the full shower phase space is to study classification performance, as described in the next section.




\subsection{Classification as a Performance Proxy}
\label{sec:classification}

When training a six-layer, fully-connected classification model on the 504-dimensional pixel space of the concatenated representation of shower energy depositions across all calorimeter layers, no major classification degradation is observed for out-of-domain learning when trained on the full simulation, \textit{i.e.} when the network is trained on \textsc{Geant4} samples but evaluated on \textsc{CaloGAN} samples. Specifically, although the classification accuracy always reaches 99\% when evaluating performance on \textsc{CaloGAN} showers -- which points to an over-differentiation among particle types in the \textsc{CaloGAN} dataset -- in both $e^+ - \gamma$ and $e^+ - \pi^+$ discrimination tasks, the evaluation of the network trained on \textsc{Geant4} images results in no accuracy decrease in the former task ($\sim 70 \%$), and only a 2\% decrease in the latter ($\sim 97\%$ versus $\gtrsim 99\%$ accuracy), when compared to the classifier tested on \textsc{CaloGAN} samples. The stability of the accuracy metric implies that the \textsc{CaloGAN} succeeds at representing at least as much variation among showers initiated by different particles as it is necessary to classify them using the same features in \textsc{Geant4}. Training on \textsc{CaloGAN} and testing on \textsc{Geant4} does show significant degradation, indicating that the GAN is inventing new class-dependent features or underrepresenting class-independent features.  While percent-level variations may be important for some applications, using classification as a generator diagnostic is an important tool for exposing the modeling of interclass shower variations.


\subsection{Computational Performance}

Directly generating deposited energy per calorimeter cell rather than particle dynamics renders the model's time-complexity invariant to nominal energy, whereas \textsc{Geant4} shower simulation runtime increases significantly with higher energy.  Therefore, the \textsc{CaloGAN} affords sizable simulation-time speed ups compared to \textsc{Geant4}. All benchmarks are performed on Intel Xeon\textsuperscript{\textregistered} 2.6GHz processors for CPU-time and a single NVIDIA\textsuperscript{\textregistered} K80 for GPU-time. When simulating a single $e^{+}$ in a uniform energy range between 1 GeV and 100 GeV, \textsc{CaloGAN} is $\mathcal{O}(10^2)$ times faster than \textsc{Geant4} on both CPU and GPU. However, when batching is utilized, the \textsc{CaloGAN} throughput significantly improves -- when batching of size 1024 is allowed (not unrealistic given the embarrassingly parallel nature of EM showering), the per-$e^{+}$ generation time is $\mathcal{O}(10^3)$ times faster on CPU and $\mathcal{O}(10^5)$ times faster on GPU.

\section{Outlook and Future Work}
This letter demonstrates that the Generative Adversarial Network technology represents a powerful new tool for efficient simulation. Our ability to infuse Physics domain knowledge into the neural network documents the flexibility and extensibility of the method for field-specific applications and explicit mismodeling mitigation. 

Prior to this work, the prospect of a GAN-based calorimeter simulation had generated considerable excitement within the high energy physics community. The availability and performance of the \textsc{CaloGAN} has attracted further interest as a concrete and publically available demonstration of the power and drawbacks of a GAN-based calorimeter simulation. In addition to the applicability within individual experiments, variations of the \textsc{CaloGAN} are also being studied as a generic tool for future \textsc{Geant} software versions. While the \textsc{CaloGAN} is currently structured as a fast simulation tool, in the future it could also be trained on testbeam data to replace or augment a full simulation tool.


Future work will focus on incorporating the most recent cutting-edge innovations from the GAN literature to stabilize the training procedure and improve convergence to optimal solutions~\cite{improved_wgan,fgan,ttugan,began}.
While our primary effort will be to improve and maintain this technique for event simulation at the LHC, this neural-network approach retains generalization power to other fields in which computationally expensive simulation inhibits result productivity.

\section{acknowledgments}
This work was supported in part by the Office of High Energy Physics of the U.S. Department of Energy under contracts DE-AC02-05CH11231 and DE-FG02-92ER40704. The authors would like to thank Wahid Bhimji, Zach Marshall, Mustafa Mustafa, and Prabhat, for helpful conversations.

\bibliography{refs}

\begin{thebibliography}{36}%
\makeatletter
\providecommand \@ifxundefined [1]{%
 \@ifx{#1\undefined}
}%
\providecommand \@ifnum [1]{%
 \ifnum #1\expandafter \@firstoftwo
 \else \expandafter \@secondoftwo
 \fi
}%
\providecommand \@ifx [1]{%
 \ifx #1\expandafter \@firstoftwo
 \else \expandafter \@secondoftwo
 \fi
}%
\providecommand \natexlab [1]{#1}%
\providecommand \enquote  [1]{``#1''}%
\providecommand \bibnamefont  [1]{#1}%
\providecommand \bibfnamefont [1]{#1}%
\providecommand \citenamefont [1]{#1}%
\providecommand \href@noop [0]{\@secondoftwo}%
\providecommand \href [0]{\begingroup \@sanitize@url \@href}%
\providecommand \@href[1]{\@@startlink{#1}\@@href}%
\providecommand \@@href[1]{\endgroup#1\@@endlink}%
\providecommand \@sanitize@url [0]{\catcode `\\12\catcode `\$12\catcode
  `\&12\catcode `\#12\catcode `\^12\catcode `\_12\catcode `\%12\relax}%
\providecommand \@@startlink[1]{}%
\providecommand \@@endlink[0]{}%
\providecommand \url  [0]{\begingroup\@sanitize@url \@url }%
\providecommand \@url [1]{\endgroup\@href {#1}{\urlprefix }}%
\providecommand \urlprefix  [0]{URL }%
\providecommand \Eprint [0]{\href }%
\providecommand \doibase [0]{http://dx.doi.org/}%
\providecommand \selectlanguage [0]{\@gobble}%
\providecommand \bibinfo  [0]{\@secondoftwo}%
\providecommand \bibfield  [0]{\@secondoftwo}%
\providecommand \translation [1]{[#1]}%
\providecommand \BibitemOpen [0]{}%
\providecommand \bibitemStop [0]{}%
\providecommand \bibitemNoStop [0]{.\EOS\space}%
\providecommand \EOS [0]{\spacefactor3000\relax}%
\providecommand \BibitemShut  [1]{\csname bibitem#1\endcsname}%
\let\auto@bib@innerbib\@empty
\bibitem [{\citenamefont {Flynn}(2015)}]{Flynn:2002240}%
  \BibitemOpen
  \bibfield  {author} {\bibinfo {author} {\bibfnamefont {J.}~\bibnamefont
  {Flynn}},\ }\href {https://cds.cern.ch/record/2002240} {\emph {\bibinfo
  {title} {{Computing Resources Scrutiny Group Report}}}},\ \bibinfo {type}
  {Tech. Rep.}\ \bibinfo {number} {CERN-RRB-2015-014}\ (\bibinfo  {institution}
  {CERN},\ \bibinfo {address} {Geneva},\ \bibinfo {year} {2015})\BibitemShut
  {NoStop}%
\bibitem [{\citenamefont {Karavakis}\ \emph {et~al.}(2014)\citenamefont
  {Karavakis} \emph {et~al.}}]{dashboard}%
  \BibitemOpen
  \bibfield  {author} {\bibinfo {author} {\bibfnamefont {E.}~\bibnamefont
  {Karavakis}} \emph {et~al.},\ }\href
  {http://stacks.iop.org/1742-6596/513/i=6/a=062024} {\bibfield  {journal}
  {\bibinfo  {journal} {Journal of Physics: Conference Series}\ }\textbf
  {\bibinfo {volume} {513}},\ \bibinfo {pages} {062024} (\bibinfo {year}
  {2014})}\BibitemShut {NoStop}%
\bibitem [{\citenamefont {Bozzi}(2015)}]{Bozzi:1984010}%
  \BibitemOpen
  \bibfield  {author} {\bibinfo {author} {\bibfnamefont {C.}~\bibnamefont
  {Bozzi}},\ }\href {https://cds.cern.ch/record/1984010} {\emph {\bibinfo
  {title} {{LHCb Computing Resource usage in 2014 (II)}}}},\ \bibinfo {type}
  {Tech. Rep.}\ \bibinfo {number} {LHCb-PUB-2015-004. CERN-LHCb-PUB-2015-004}\
  (\bibinfo  {institution} {CERN},\ \bibinfo {address} {Geneva},\ \bibinfo
  {year} {2015})\BibitemShut {NoStop}%
\bibitem [{\citenamefont {Aad}\ \emph {et~al.}(2012)\citenamefont {Aad} \emph
  {et~al.}}]{Aad:2012tfa}%
  \BibitemOpen
  \bibfield  {author} {\bibinfo {author} {\bibfnamefont {G.}~\bibnamefont
  {Aad}} \emph {et~al.} (\bibinfo {collaboration} {ATLAS}),\ }\href {\doibase
  10.1016/j.physletb.2012.08.020} {\bibfield  {journal} {\bibinfo  {journal}
  {Phys. Lett.}\ }\textbf {\bibinfo {volume} {B716}},\ \bibinfo {pages} {1}
  (\bibinfo {year} {2012})},\ \Eprint {http://arxiv.org/abs/1207.7214}
  {arXiv:1207.7214 [hep-ex]} \BibitemShut {NoStop}%
\bibitem [{\citenamefont {Chatrchyan}\ \emph {et~al.}(2012)\citenamefont
  {Chatrchyan} \emph {et~al.}}]{Chatrchyan:2012xdj}%
  \BibitemOpen
  \bibfield  {author} {\bibinfo {author} {\bibfnamefont {S.}~\bibnamefont
  {Chatrchyan}} \emph {et~al.} (\bibinfo {collaboration} {CMS}),\ }\href
  {\doibase 10.1016/j.physletb.2012.08.021} {\bibfield  {journal} {\bibinfo
  {journal} {Phys. Lett.}\ }\textbf {\bibinfo {volume} {B716}},\ \bibinfo
  {pages} {30} (\bibinfo {year} {2012})},\ \Eprint
  {http://arxiv.org/abs/1207.7235} {arXiv:1207.7235 [hep-ex]} \BibitemShut
  {NoStop}%
\bibitem [{\citenamefont {{GEANT4 Collaboration}}(2003)}]{Geant}%
  \BibitemOpen
  \bibfield  {author} {\bibinfo {author} {\bibnamefont {{GEANT4
  Collaboration}}},\ }\href@noop {} {\bibfield  {journal} {\bibinfo  {journal}
  {Nuclear Instruments and Methods in Physics Research A}\ }\textbf {\bibinfo
  {volume} {506}},\ \bibinfo {pages} {250} (\bibinfo {year}
  {2003})}\BibitemShut {NoStop}%
\bibitem [{\citenamefont {Aad}\ \emph {et~al.}(2010)\citenamefont {Aad} \emph
  {et~al.}}]{Aad:2010ah}%
  \BibitemOpen
  \bibfield  {author} {\bibinfo {author} {\bibfnamefont {G.}~\bibnamefont
  {Aad}} \emph {et~al.} (\bibinfo {collaboration} {ATLAS}),\ }\href {\doibase
  10.1140/epjc/s10052-010-1429-9} {\bibfield  {journal} {\bibinfo  {journal}
  {Eur. Phys. J.}\ }\textbf {\bibinfo {volume} {C70}},\ \bibinfo {pages} {823}
  (\bibinfo {year} {2010})},\ \Eprint {http://arxiv.org/abs/1005.4568}
  {arXiv:1005.4568 [physics.ins-det]} \BibitemShut {NoStop}%
\bibitem [{\citenamefont {Rahmat}\ \emph {et~al.}(2012)\citenamefont {Rahmat},
  \citenamefont {Kroeger},\ and\ \citenamefont
  {Giammanco}}]{1742-6596-396-6-062016}%
  \BibitemOpen
  \bibfield  {author} {\bibinfo {author} {\bibfnamefont {R.}~\bibnamefont
  {Rahmat}}, \bibinfo {author} {\bibfnamefont {R.}~\bibnamefont {Kroeger}}, \
  and\ \bibinfo {author} {\bibfnamefont {A.}~\bibnamefont {Giammanco}},\ }\href
  {http://stacks.iop.org/1742-6596/396/i=6/a=062016} {\bibfield  {journal}
  {\bibinfo  {journal} {Journal of Physics: Conference Series}\ }\textbf
  {\bibinfo {volume} {396}},\ \bibinfo {pages} {062016} (\bibinfo {year}
  {2012})}\BibitemShut {NoStop}%
\bibitem [{\citenamefont {de~Florian}\ \emph {et~al.}(2016)\citenamefont
  {de~Florian} \emph {et~al.}}]{deFlorian:2016spz}%
  \BibitemOpen
  \bibfield  {author} {\bibinfo {author} {\bibfnamefont {D.}~\bibnamefont
  {de~Florian}} \emph {et~al.} (\bibinfo {collaboration} {LHC Higgs Cross
  Section Working Group}),\ }\href {\doibase 10.23731/CYRM-2017-002} {\
  (\bibinfo {year} {2016}),\ 10.23731/CYRM-2017-002},\ \Eprint
  {http://arxiv.org/abs/1610.07922} {arXiv:1610.07922 [hep-ph]} \BibitemShut
  {NoStop}%
\bibitem [{\citenamefont {Aaboud}\ \emph {et~al.}(2016)\citenamefont {Aaboud}
  \emph {et~al.}}]{Aaboud:2016mmw}%
  \BibitemOpen
  \bibfield  {author} {\bibinfo {author} {\bibfnamefont {M.}~\bibnamefont
  {Aaboud}} \emph {et~al.} (\bibinfo {collaboration} {ATLAS}),\ }\href
  {\doibase 10.1103/PhysRevLett.117.182002} {\bibfield  {journal} {\bibinfo
  {journal} {Phys. Rev. Lett.}\ }\textbf {\bibinfo {volume} {117}},\ \bibinfo
  {pages} {182002} (\bibinfo {year} {2016})},\ \Eprint
  {http://arxiv.org/abs/1606.02625} {arXiv:1606.02625 [hep-ex]} \BibitemShut
  {NoStop}%
\bibitem [{CMS(2016)}]{CMS-PAS-FSQ-15-005}%
  \BibitemOpen
  \href {https://cds.cern.ch/record/2145896} {\emph {\bibinfo {title}
  {{Measurement of the inelastic proton-proton cross section at
  $\sqrt{s}=13~\mathrm{TeV}$}}}},\ \bibinfo {type} {Tech. Rep.}\ \bibinfo
  {number} {CMS-PAS-FSQ-15-005}\ (\bibinfo  {institution} {CERN},\ \bibinfo
  {address} {Geneva},\ \bibinfo {year} {2016})\BibitemShut {NoStop}%
\bibitem [{\citenamefont {Grindhammer}\ and\ \citenamefont
  {Peters}(1993)}]{Grindhammer:1993kw}%
  \BibitemOpen
  \bibfield  {author} {\bibinfo {author} {\bibfnamefont {G.}~\bibnamefont
  {Grindhammer}}\ and\ \bibinfo {author} {\bibfnamefont {S.}~\bibnamefont
  {Peters}},\ }in\ \href {http://alice.cern.ch/format/showfull?sysnb=2168235}
  {\emph {\bibinfo {booktitle} {{International Conference on Monte Carlo
  Simulation in High-Energy and Nuclear Physics - MC 93 Tallahassee, Florida,
  February 22-26, 1993}}}}\ (\bibinfo {year} {1993})\ \Eprint
  {http://arxiv.org/abs/hep-ex/0001020} {arXiv:hep-ex/0001020 [hep-ex]}
  \BibitemShut {NoStop}%
\bibitem [{\citenamefont {Beckingham}\ \emph {et~al.}(2010)\citenamefont
  {Beckingham} \emph {et~al.}}]{ATLAS:1300517}%
  \BibitemOpen
  \bibfield  {author} {\bibinfo {author} {\bibfnamefont {M.}~\bibnamefont
  {Beckingham}} \emph {et~al.} (\bibinfo {collaboration} {{ATLAS}}),\ }\href
  {https://cds.cern.ch/record/1300517} {\emph {\bibinfo {title} {{The
  simulation principle and performance of the ATLAS fast calorimeter simulation
  FastCaloSim}}}},\ \bibinfo {type} {Tech. Rep.}\ \bibinfo {number}
  {ATL-PHYS-PUB-2010-013}\ (\bibinfo  {institution} {CERN},\ \bibinfo {address}
  {Geneva},\ \bibinfo {year} {2010})\BibitemShut {NoStop}%
\bibitem [{\citenamefont {Grindhammer}\ \emph {et~al.}(1990)\citenamefont
  {Grindhammer}, \citenamefont {Rudowicz},\ and\ \citenamefont
  {Peters}}]{Grindhammer:1989zg}%
  \BibitemOpen
  \bibfield  {author} {\bibinfo {author} {\bibfnamefont {G.}~\bibnamefont
  {Grindhammer}}, \bibinfo {author} {\bibfnamefont {M.}~\bibnamefont
  {Rudowicz}}, \ and\ \bibinfo {author} {\bibfnamefont {S.}~\bibnamefont
  {Peters}},\ }\bibfield  {booktitle} {\emph {\bibinfo {booktitle} {{SSC
  Workshop on Calorimetry for the Superconducting Super Collider Tuscaloosa,
  Alabama, March 13-17, 1989}}},\ }\href {\doibase
  10.1016/0168-9002(90)90566-O} {\bibfield  {journal} {\bibinfo  {journal}
  {Nucl. Instrum. Meth.}\ }\textbf {\bibinfo {volume} {A290}},\ \bibinfo
  {pages} {469} (\bibinfo {year} {1990})}\BibitemShut {NoStop}%
\bibitem [{\citenamefont {Barberio}\ \emph {et~al.}(2009)\citenamefont
  {Barberio}, \citenamefont {Boudreau}, \citenamefont {Butler}, \citenamefont
  {Cheung}, \citenamefont {Dell'Acqua}, \citenamefont {Simone}, \citenamefont
  {Ehrenfeld}, \citenamefont {Gallas}, \citenamefont {Glazov}, \citenamefont
  {Marshall}, \citenamefont {Mueller}, \citenamefont {PlaÄakyte},
  \citenamefont {Rimoldi}, \citenamefont {Savard}, \citenamefont {Tsulaia},
  \citenamefont {Waugh},\ and\ \citenamefont {Young}}]{frozen}%
  \BibitemOpen
  \bibfield  {author} {\bibinfo {author} {\bibfnamefont {E.}~\bibnamefont
  {Barberio}}, \bibinfo {author} {\bibfnamefont {J.}~\bibnamefont {Boudreau}},
  \bibinfo {author} {\bibfnamefont {B.}~\bibnamefont {Butler}}, \bibinfo
  {author} {\bibfnamefont {S.~L.}\ \bibnamefont {Cheung}}, \bibinfo {author}
  {\bibfnamefont {A.}~\bibnamefont {Dell'Acqua}}, \bibinfo {author}
  {\bibfnamefont {A.~D.}\ \bibnamefont {Simone}}, \bibinfo {author}
  {\bibfnamefont {E.}~\bibnamefont {Ehrenfeld}}, \bibinfo {author}
  {\bibfnamefont {M.~V.}\ \bibnamefont {Gallas}}, \bibinfo {author}
  {\bibfnamefont {A.}~\bibnamefont {Glazov}}, \bibinfo {author} {\bibfnamefont
  {Z.}~\bibnamefont {Marshall}}, \bibinfo {author} {\bibfnamefont
  {J.}~\bibnamefont {Mueller}}, \bibinfo {author} {\bibfnamefont
  {R.}~\bibnamefont {PlaÄakyte}}, \bibinfo {author} {\bibfnamefont
  {A.}~\bibnamefont {Rimoldi}}, \bibinfo {author} {\bibfnamefont
  {P.}~\bibnamefont {Savard}}, \bibinfo {author} {\bibfnamefont
  {V.}~\bibnamefont {Tsulaia}}, \bibinfo {author} {\bibfnamefont
  {A.}~\bibnamefont {Waugh}}, \ and\ \bibinfo {author} {\bibfnamefont {C.~C.}\
  \bibnamefont {Young}},\ }\href
  {http://stacks.iop.org/1742-6596/160/i=1/a=012082} {\bibfield  {journal}
  {\bibinfo  {journal} {Journal of Physics: Conference Series}\ }\textbf
  {\bibinfo {volume} {160}},\ \bibinfo {pages} {012082} (\bibinfo {year}
  {2009})}\BibitemShut {NoStop}%
\bibitem [{\citenamefont {Ravanbakhsh}\ \emph {et~al.}(2016)\citenamefont
  {Ravanbakhsh}, \citenamefont {Lanusse}, \citenamefont {Mandelbaum},
  \citenamefont {Schneider},\ and\ \citenamefont {Poczos}}]{CosmoGAN1}%
  \BibitemOpen
  \bibfield  {author} {\bibinfo {author} {\bibfnamefont {S.}~\bibnamefont
  {Ravanbakhsh}}, \bibinfo {author} {\bibfnamefont {F.}~\bibnamefont
  {Lanusse}}, \bibinfo {author} {\bibfnamefont {R.}~\bibnamefont {Mandelbaum}},
  \bibinfo {author} {\bibfnamefont {J.}~\bibnamefont {Schneider}}, \ and\
  \bibinfo {author} {\bibfnamefont {B.}~\bibnamefont {Poczos}},\ }\href@noop {}
  {\  (\bibinfo {year} {2016})},\ \Eprint {http://arxiv.org/abs/1609.05796}
  {arXiv:1609.05796 [astro-ph.IM]} \BibitemShut {NoStop}%
\bibitem [{\citenamefont {{Schawinski}}\ \emph {et~al.}(2017)\citenamefont
  {{Schawinski}}, \citenamefont {{Zhang}}, \citenamefont {{Zhang}},
  \citenamefont {{Fowler}},\ and\ \citenamefont {{Santhanam}}}]{CosmoGAN2}%
  \BibitemOpen
  \bibfield  {author} {\bibinfo {author} {\bibfnamefont {K.}~\bibnamefont
  {{Schawinski}}}, \bibinfo {author} {\bibfnamefont {C.}~\bibnamefont
  {{Zhang}}}, \bibinfo {author} {\bibfnamefont {H.}~\bibnamefont {{Zhang}}},
  \bibinfo {author} {\bibfnamefont {L.}~\bibnamefont {{Fowler}}}, \ and\
  \bibinfo {author} {\bibfnamefont {G.~K.}\ \bibnamefont {{Santhanam}}},\
  }\href@noop {} {\ \textbf {\bibinfo {volume} {467}} (\bibinfo {year}
  {2017})},\ \Eprint {http://arxiv.org/abs/1702.00403} {arXiv:1702.00403
  [astro-ph.IM]} \BibitemShut {NoStop}%
\bibitem [{\citenamefont {{Mosser}}\ \emph {et~al.}(2017)\citenamefont
  {{Mosser}}, \citenamefont {{Dubrule}},\ and\ \citenamefont
  {{Blunt}}}]{condmatter}%
  \BibitemOpen
  \bibfield  {author} {\bibinfo {author} {\bibfnamefont {L.}~\bibnamefont
  {{Mosser}}}, \bibinfo {author} {\bibfnamefont {O.}~\bibnamefont {{Dubrule}}},
  \ and\ \bibinfo {author} {\bibfnamefont {M.~J.}\ \bibnamefont {{Blunt}}},\
  }\href@noop {} {\bibfield  {journal} {\bibinfo  {journal} {ArXiv e-prints}\ }
  (\bibinfo {year} {2017})},\ \Eprint {http://arxiv.org/abs/1704.03225}
  {arXiv:1704.03225 [cs.CV]} \BibitemShut {NoStop}%
\bibitem [{\citenamefont {Kadurin}\ \emph {et~al.}(2016)\citenamefont
  {Kadurin}, \citenamefont {Aliper}, \citenamefont {Kazennov}, \citenamefont
  {Mamoshina}, \citenamefont {Vanhaelen}, \citenamefont {Khrabrov},\ and\
  \citenamefont {Zhavoronkov}}]{oncology}%
  \BibitemOpen
  \bibfield  {author} {\bibinfo {author} {\bibfnamefont {A.}~\bibnamefont
  {Kadurin}}, \bibinfo {author} {\bibfnamefont {A.}~\bibnamefont {Aliper}},
  \bibinfo {author} {\bibfnamefont {A.}~\bibnamefont {Kazennov}}, \bibinfo
  {author} {\bibfnamefont {P.}~\bibnamefont {Mamoshina}}, \bibinfo {author}
  {\bibfnamefont {Q.}~\bibnamefont {Vanhaelen}}, \bibinfo {author}
  {\bibfnamefont {K.}~\bibnamefont {Khrabrov}}, \ and\ \bibinfo {author}
  {\bibfnamefont {A.}~\bibnamefont {Zhavoronkov}},\ }\href
  {http://www.impactjournals.com/oncotarget/index.php?journal=oncotarget&page=article&op=view&path%5B%5D=14073}
  {\bibfield  {journal} {\bibinfo  {journal} {Oncotarget}\ }\textbf {\bibinfo
  {volume} {8}} (\bibinfo {year} {2016})}\BibitemShut {NoStop}%
\bibitem [{\citenamefont {Goodfellow}\ \emph {et~al.}(2014)\citenamefont
  {Goodfellow}, \citenamefont {Pouget-Abadie}, \citenamefont {Mirza},
  \citenamefont {Xu}, \citenamefont {Warde-Farley}, \citenamefont {Ozair},
  \citenamefont {Courville},\ and\ \citenamefont
  {Bengio}}]{goodfellow2014generative}%
  \BibitemOpen
  \bibfield  {author} {\bibinfo {author} {\bibfnamefont {I.~J.}\ \bibnamefont
  {Goodfellow}}, \bibinfo {author} {\bibfnamefont {J.}~\bibnamefont
  {Pouget-Abadie}}, \bibinfo {author} {\bibfnamefont {M.}~\bibnamefont
  {Mirza}}, \bibinfo {author} {\bibfnamefont {B.}~\bibnamefont {Xu}}, \bibinfo
  {author} {\bibfnamefont {D.}~\bibnamefont {Warde-Farley}}, \bibinfo {author}
  {\bibfnamefont {S.}~\bibnamefont {Ozair}}, \bibinfo {author} {\bibfnamefont
  {A.}~\bibnamefont {Courville}}, \ and\ \bibinfo {author} {\bibfnamefont
  {Y.}~\bibnamefont {Bengio}},\ }\href@noop {} {\bibfield  {journal} {\bibinfo
  {journal} {ArXiv e-prints}\ } (\bibinfo {year} {2014})},\ \Eprint
  {http://arxiv.org/abs/1406.2661} {arXiv:1406.2661} \BibitemShut {NoStop}%
\bibitem [{\citenamefont {Radford}\ \emph {et~al.}(2015)\citenamefont
  {Radford}, \citenamefont {Metz},\ and\ \citenamefont {Chintala}}]{dcgan}%
  \BibitemOpen
  \bibfield  {author} {\bibinfo {author} {\bibfnamefont {A.}~\bibnamefont
  {Radford}}, \bibinfo {author} {\bibfnamefont {L.}~\bibnamefont {Metz}}, \
  and\ \bibinfo {author} {\bibfnamefont {S.}~\bibnamefont {Chintala}},\ }\href
  {http://arxiv.org/abs/1511.06434} {\bibfield  {journal} {\bibinfo  {journal}
  {ArXiv e-prints}\ } (\bibinfo {year} {2015})},\ \Eprint
  {http://arxiv.org/abs/1511.06434} {arXiv:1511.06434} \BibitemShut {NoStop}%
\bibitem [{\citenamefont {Chetlur}\ \emph {et~al.}(2014)\citenamefont
  {Chetlur}, \citenamefont {Woolley}, \citenamefont {Vandermersch},
  \citenamefont {Cohen}, \citenamefont {Tran}, \citenamefont {Catanzaro},\ and\
  \citenamefont {Shelhamer}}]{DBLP:journals/corr/ChetlurWVCTCS14}%
  \BibitemOpen
  \bibfield  {author} {\bibinfo {author} {\bibfnamefont {S.}~\bibnamefont
  {Chetlur}}, \bibinfo {author} {\bibfnamefont {C.}~\bibnamefont {Woolley}},
  \bibinfo {author} {\bibfnamefont {P.}~\bibnamefont {Vandermersch}}, \bibinfo
  {author} {\bibfnamefont {J.}~\bibnamefont {Cohen}}, \bibinfo {author}
  {\bibfnamefont {J.}~\bibnamefont {Tran}}, \bibinfo {author} {\bibfnamefont
  {B.}~\bibnamefont {Catanzaro}}, \ and\ \bibinfo {author} {\bibfnamefont
  {E.}~\bibnamefont {Shelhamer}},\ }\href {http://arxiv.org/abs/1410.0759}
  {\bibfield  {journal} {\bibinfo  {journal} {ArXiv e-prints}\ } (\bibinfo
  {year} {2014})},\ \Eprint {http://arxiv.org/abs/1410.0759} {arXiv:1410.0759
  [cs.NE]} \BibitemShut {NoStop}%
\bibitem [{\citenamefont {de~Oliveira}\ \emph {et~al.}(2017)\citenamefont
  {de~Oliveira}, \citenamefont {Paganini},\ and\ \citenamefont
  {Nachman}}]{deOliveira:2017pjk}%
  \BibitemOpen
  \bibfield  {author} {\bibinfo {author} {\bibfnamefont {L.}~\bibnamefont
  {de~Oliveira}}, \bibinfo {author} {\bibfnamefont {M.}~\bibnamefont
  {Paganini}}, \ and\ \bibinfo {author} {\bibfnamefont {B.}~\bibnamefont
  {Nachman}},\ }\href@noop {} {\  (\bibinfo {year} {2017})},\ \Eprint
  {http://arxiv.org/abs/1701.05927} {arXiv:1701.05927 [stat.ML]} \BibitemShut
  {NoStop}%
\bibitem [{\citenamefont {Cogan}\ \emph {et~al.}(2015)\citenamefont {Cogan},
  \citenamefont {Kagan}, \citenamefont {Strauss},\ and\ \citenamefont
  {Schwarztman}}]{Cogan:2014oua}%
  \BibitemOpen
  \bibfield  {author} {\bibinfo {author} {\bibfnamefont {J.}~\bibnamefont
  {Cogan}}, \bibinfo {author} {\bibfnamefont {M.}~\bibnamefont {Kagan}},
  \bibinfo {author} {\bibfnamefont {E.}~\bibnamefont {Strauss}}, \ and\
  \bibinfo {author} {\bibfnamefont {A.}~\bibnamefont {Schwarztman}},\ }\href
  {\doibase 10.1007/JHEP02(2015)118} {\bibfield  {journal} {\bibinfo  {journal}
  {JHEP}\ }\textbf {\bibinfo {volume} {02}},\ \bibinfo {pages} {118} (\bibinfo
  {year} {2015})},\ \Eprint {http://arxiv.org/abs/1407.5675} {arXiv:1407.5675
  [hep-ph]} \BibitemShut {NoStop}%
\bibitem [{\citenamefont {Sjostrand}\ \emph {et~al.}(2006)\citenamefont
  {Sjostrand}, \citenamefont {Mrenna},\ and\ \citenamefont
  {Skands}}]{Sjostrand:2006za}%
  \BibitemOpen
  \bibfield  {author} {\bibinfo {author} {\bibfnamefont {T.}~\bibnamefont
  {Sjostrand}}, \bibinfo {author} {\bibfnamefont {S.}~\bibnamefont {Mrenna}}, \
  and\ \bibinfo {author} {\bibfnamefont {P.~Z.}\ \bibnamefont {Skands}},\
  }\href {\doibase 10.1088/1126-6708/2006/05/026} {\bibfield  {journal}
  {\bibinfo  {journal} {JHEP}\ }\textbf {\bibinfo {volume} {05}},\ \bibinfo
  {pages} {026} (\bibinfo {year} {2006})},\ \Eprint
  {http://arxiv.org/abs/hep-ph/0603175} {arXiv:hep-ph/0603175 [hep-ph]}
  \BibitemShut {NoStop}%
\bibitem [{\citenamefont {Xu}\ \emph {et~al.}(2015)\citenamefont {Xu},
  \citenamefont {Ba}, \citenamefont {Kiros}, \citenamefont {Cho}, \citenamefont
  {Courville}, \citenamefont {Salakhudinov}, \citenamefont {Zemel},\ and\
  \citenamefont {Bengio}}]{attention}%
  \BibitemOpen
  \bibfield  {author} {\bibinfo {author} {\bibfnamefont {K.}~\bibnamefont
  {Xu}}, \bibinfo {author} {\bibfnamefont {J.}~\bibnamefont {Ba}}, \bibinfo
  {author} {\bibfnamefont {R.}~\bibnamefont {Kiros}}, \bibinfo {author}
  {\bibfnamefont {K.}~\bibnamefont {Cho}}, \bibinfo {author} {\bibfnamefont
  {A.}~\bibnamefont {Courville}}, \bibinfo {author} {\bibfnamefont
  {R.}~\bibnamefont {Salakhudinov}}, \bibinfo {author} {\bibfnamefont
  {R.}~\bibnamefont {Zemel}}, \ and\ \bibinfo {author} {\bibfnamefont
  {Y.}~\bibnamefont {Bengio}},\ }in\ \href
  {http://proceedings.mlr.press/v37/xuc15.html} {\emph {\bibinfo {booktitle}
  {Proceedings of the 32nd International Conference on Machine Learning}}},\
  \bibinfo {series} {Proceedings of Machine Learning Research}, Vol.~\bibinfo
  {volume} {37}\ (\bibinfo  {publisher} {PMLR},\ \bibinfo {address} {Lille,
  France},\ \bibinfo {year} {2015})\ pp.\ \bibinfo {pages}
  {2048--2057}\BibitemShut {NoStop}%
\bibitem [{\citenamefont {{Zhang}}\ \emph {et~al.}(2016)\citenamefont
  {{Zhang}}, \citenamefont {{Xu}}, \citenamefont {{Li}}, \citenamefont
  {{Zhang}}, \citenamefont {{Huang}}, \citenamefont {{Wang}},\ and\
  \citenamefont {{Metaxas}}}]{stackgan}%
  \BibitemOpen
  \bibfield  {author} {\bibinfo {author} {\bibfnamefont {H.}~\bibnamefont
  {{Zhang}}}, \bibinfo {author} {\bibfnamefont {T.}~\bibnamefont {{Xu}}},
  \bibinfo {author} {\bibfnamefont {H.}~\bibnamefont {{Li}}}, \bibinfo {author}
  {\bibfnamefont {S.}~\bibnamefont {{Zhang}}}, \bibinfo {author} {\bibfnamefont
  {X.}~\bibnamefont {{Huang}}}, \bibinfo {author} {\bibfnamefont
  {X.}~\bibnamefont {{Wang}}}, \ and\ \bibinfo {author} {\bibfnamefont
  {D.}~\bibnamefont {{Metaxas}}},\ }\href@noop {} {\bibfield  {journal}
  {\bibinfo  {journal} {ArXiv e-prints}\ } (\bibinfo {year} {2016})},\ \Eprint
  {http://arxiv.org/abs/1612.03242} {arXiv:1612.03242 [cs.CV]} \BibitemShut
  {NoStop}%
\bibitem [{\citenamefont {Salimans}\ \emph {et~al.}(2016)\citenamefont
  {Salimans}, \citenamefont {Goodfellow}, \citenamefont {Zaremba},
  \citenamefont {Cheung}, \citenamefont {Radford},\ and\ \citenamefont
  {Chen}}]{improved_gan}%
  \BibitemOpen
  \bibfield  {author} {\bibinfo {author} {\bibfnamefont {T.}~\bibnamefont
  {Salimans}}, \bibinfo {author} {\bibfnamefont {I.~J.}\ \bibnamefont
  {Goodfellow}}, \bibinfo {author} {\bibfnamefont {W.}~\bibnamefont {Zaremba}},
  \bibinfo {author} {\bibfnamefont {V.}~\bibnamefont {Cheung}}, \bibinfo
  {author} {\bibfnamefont {A.}~\bibnamefont {Radford}}, \ and\ \bibinfo
  {author} {\bibfnamefont {X.}~\bibnamefont {Chen}},\ }\href
  {http://arxiv.org/abs/1606.03498} {\bibfield  {journal} {\bibinfo  {journal}
  {ArXiv e-prints}\ } (\bibinfo {year} {2016})},\ \Eprint
  {http://arxiv.org/abs/1606.03498} {arXiv:1606.03498} \BibitemShut {NoStop}%
\bibitem [{\citenamefont {Nachman}\ \emph
  {et~al.}(2017{\natexlab{a}})\citenamefont {Nachman}, \citenamefont
  {de~Oliveira},\ and\ \citenamefont {Paganini}}]{dataset}%
  \BibitemOpen
  \bibfield  {author} {\bibinfo {author} {\bibfnamefont {B.}~\bibnamefont
  {Nachman}}, \bibinfo {author} {\bibfnamefont {L.}~\bibnamefont
  {de~Oliveira}}, \ and\ \bibinfo {author} {\bibfnamefont {M.}~\bibnamefont
  {Paganini}},\ }\href {\doibase 10.17632/pvn3xc3wy5.1} {\  (\bibinfo {year}
  {2017}{\natexlab{a}}),\ 10.17632/pvn3xc3wy5.1}\BibitemShut {NoStop}%
\bibitem [{\citenamefont {Aad}\ \emph {et~al.}(2017)\citenamefont {Aad} \emph
  {et~al.}}]{Aad:2016upy}%
  \BibitemOpen
  \bibfield  {author} {\bibinfo {author} {\bibfnamefont {G.}~\bibnamefont
  {Aad}} \emph {et~al.} (\bibinfo {collaboration} {ATLAS}),\ }\href {\doibase
  10.1140/epjc/s10052-017-5004-5} {\bibfield  {journal} {\bibinfo  {journal}
  {Eur. Phys. J.}\ }\textbf {\bibinfo {volume} {C77}},\ \bibinfo {pages} {490}
  (\bibinfo {year} {2017})},\ \Eprint {http://arxiv.org/abs/1603.02934}
  {arXiv:1603.02934 [hep-ex]} \BibitemShut {NoStop}%
\bibitem [{\citenamefont {Nachman}\ \emph
  {et~al.}(2017{\natexlab{b}})\citenamefont {Nachman}, \citenamefont
  {de~Oliveira},\ and\ \citenamefont {Paganini}}]{code_new}%
  \BibitemOpen
  \bibfield  {author} {\bibinfo {author} {\bibfnamefont {B.}~\bibnamefont
  {Nachman}}, \bibinfo {author} {\bibfnamefont {L.}~\bibnamefont
  {de~Oliveira}}, \ and\ \bibinfo {author} {\bibfnamefont {M.}~\bibnamefont
  {Paganini}},\ }\href {\doibase 10.5281/zenodo.584155} {\  (\bibinfo {year}
  {2017}{\natexlab{b}}),\ 10.5281/zenodo.584155}\BibitemShut {NoStop}%
\bibitem [{\citenamefont {Olive}\ \emph {et~al.}(2014)\citenamefont {Olive}
  \emph {et~al.}}]{Agashe:2014kda}%
  \BibitemOpen
  \bibfield  {author} {\bibinfo {author} {\bibfnamefont {K.~A.}\ \bibnamefont
  {Olive}} \emph {et~al.} (\bibinfo {collaboration} {Particle Data Group}),\
  }\href {\doibase 10.1088/1674-1137/38/9/090001} {\bibfield  {journal}
  {\bibinfo  {journal} {Chin. Phys.}\ }\textbf {\bibinfo {volume} {C38}},\
  \bibinfo {pages} {090001} (\bibinfo {year} {2014})}\BibitemShut {NoStop}%
\bibitem [{\citenamefont {{Gulrajani}}\ \emph {et~al.}(2017)\citenamefont
  {{Gulrajani}}, \citenamefont {{Ahmed}}, \citenamefont {{Arjovsky}},
  \citenamefont {{Dumoulin}},\ and\ \citenamefont
  {{Courville}}}]{improved_wgan}%
  \BibitemOpen
  \bibfield  {author} {\bibinfo {author} {\bibfnamefont {I.}~\bibnamefont
  {{Gulrajani}}}, \bibinfo {author} {\bibfnamefont {F.}~\bibnamefont
  {{Ahmed}}}, \bibinfo {author} {\bibfnamefont {M.}~\bibnamefont {{Arjovsky}}},
  \bibinfo {author} {\bibfnamefont {V.}~\bibnamefont {{Dumoulin}}}, \ and\
  \bibinfo {author} {\bibfnamefont {A.}~\bibnamefont {{Courville}}},\
  }\href@noop {} {\bibfield  {journal} {\bibinfo  {journal} {ArXiv e-prints}\ }
  (\bibinfo {year} {2017})},\ \Eprint {http://arxiv.org/abs/1704.00028}
  {arXiv:1704.00028 [cs.LG]} \BibitemShut {NoStop}%
\bibitem [{\citenamefont {{Nowozin}}\ \emph {et~al.}(2016)\citenamefont
  {{Nowozin}}, \citenamefont {{Cseke}},\ and\ \citenamefont
  {{Tomioka}}}]{fgan}%
  \BibitemOpen
  \bibfield  {author} {\bibinfo {author} {\bibfnamefont {S.}~\bibnamefont
  {{Nowozin}}}, \bibinfo {author} {\bibfnamefont {B.}~\bibnamefont {{Cseke}}},
  \ and\ \bibinfo {author} {\bibfnamefont {R.}~\bibnamefont {{Tomioka}}},\
  }\href@noop {} {\bibfield  {journal} {\bibinfo  {journal} {ArXiv e-prints}\ }
  (\bibinfo {year} {2016})},\ \Eprint {http://arxiv.org/abs/1606.00709}
  {arXiv:1606.00709 [stat.ML]} \BibitemShut {NoStop}%
\bibitem [{\citenamefont {{Heusel}}\ \emph {et~al.}(2017)\citenamefont
  {{Heusel}}, \citenamefont {{Ramsauer}}, \citenamefont {{Unterthiner}},
  \citenamefont {{Nessler}}, \citenamefont {{Klambauer}},\ and\ \citenamefont
  {{Hochreiter}}}]{ttugan}%
  \BibitemOpen
  \bibfield  {author} {\bibinfo {author} {\bibfnamefont {M.}~\bibnamefont
  {{Heusel}}}, \bibinfo {author} {\bibfnamefont {H.}~\bibnamefont
  {{Ramsauer}}}, \bibinfo {author} {\bibfnamefont {T.}~\bibnamefont
  {{Unterthiner}}}, \bibinfo {author} {\bibfnamefont {B.}~\bibnamefont
  {{Nessler}}}, \bibinfo {author} {\bibfnamefont {G.}~\bibnamefont
  {{Klambauer}}}, \ and\ \bibinfo {author} {\bibfnamefont {S.}~\bibnamefont
  {{Hochreiter}}},\ }\href@noop {} {\bibfield  {journal} {\bibinfo  {journal}
  {ArXiv e-prints}\ } (\bibinfo {year} {2017})},\ \Eprint
  {http://arxiv.org/abs/1706.08500} {arXiv:1706.08500 [cs.LG]} \BibitemShut
  {NoStop}%
\bibitem [{\citenamefont {Berthelot}\ \emph {et~al.}()\citenamefont
  {Berthelot}, \citenamefont {Schumm},\ and\ \citenamefont {Metz}}]{began}%
  \BibitemOpen
  \bibfield  {author} {\bibinfo {author} {\bibfnamefont {D.}~\bibnamefont
  {Berthelot}}, \bibinfo {author} {\bibfnamefont {T.}~\bibnamefont {Schumm}}, \
  and\ \bibinfo {author} {\bibfnamefont {L.}~\bibnamefont {Metz}},\ }\href
  {http://arxiv.org/abs/1703.10717} {\bibfield  {journal} {\bibinfo  {journal}
  {ArXiv e-prints}\ }}\Eprint {http://arxiv.org/abs/1703.10717}
  {arXiv:1703.10717 [stat.LG]} \BibitemShut {NoStop}%
\end{thebibliography}%
\end{document}